\documentclass[11pt]{article}

\usepackage[final]{acl}

\usepackage{times}
\usepackage{latexsym}

\usepackage[T1]{fontenc}

\usepackage[utf8]{inputenc}

\usepackage{microtype}

\usepackage{inconsolata}

\usepackage{graphicx}

\usepackage{amsmath,amsfonts}
\usepackage{graphicx}
\usepackage{textcomp}
\usepackage{tabularx}
\usepackage{xcolor,soul}
\def\BibTeX{{\rm B\kern-.05em{\sc i\kern-.025em b}\kern-.08em
    T\kern-.1667em\lower.7ex\hbox{E}\kern-.125emX}}

\usepackage{booktabs}   
\usepackage{subcaption} 
\usepackage{array}
\usepackage{ragged2e}
\usepackage{float}
\usepackage{listings}
\usepackage{xspace}
\usepackage{multirow}
\usepackage{amsthm}
\usepackage{balance}

\usepackage[skins]{tcolorbox}

\usepackage{xcolor,pifont}
\newcommand*\colourcheck[1]{%
	\expandafter\newcommand\csname #1check\endcsname{\textcolor{#1}{\ding{52}}}%
}

\usepackage{enumitem}
\usepackage[normalem]{ulem} 
\usepackage{array}
\usepackage{wrapfig}
\usepackage{amsmath,amsfonts}
\usepackage[noend]{algpseudocode}
\usepackage{graphicx}
\usepackage{textcomp}
\usepackage{float}
\usepackage{listings}
\usepackage{xspace}
\usepackage{multirow}
\usepackage{amsthm}

\usepackage{balance}
\usepackage{algorithm}
\usepackage{algpseudocode}
\usepackage{colortbl}
\usepackage{subcaption}
\usepackage[skins]{tcolorbox}
\usepackage{xcolor}
\usepackage{multicol}
\usepackage{soul}
\usepackage{framed}
\usepackage{booktabs}
\usepackage{booktabs}
\usepackage{multirow}
\usepackage{siunitx}
\usepackage{minted}
\usepackage{multicol}
\usepackage{fancyvrb}
\usepackage{tcolorbox}
\colourcheck{blue}
\colourcheck{green}
\colourcheck{red}
\definecolor{custom-blue}{rgb}{0,0,0}
\definecolor{custom-red}{rgb}{1,0,0}

\newtcolorbox{boxB}[2][]{%
  enhanced,colback=white,colframe=black,coltitle=black,
  sharp corners,
  toprule=1.0pt,
  rightrule=0.3pt,
  leftrule=0pt,
  bottomrule=0pt,
  fonttitle=\itshape\scshape\large,
  left=0pt,right=5pt,top=5pt,bottom=3pt,
  attach boxed title to top right={yshift=-0.3\baselineskip-0.4pt,xshift=-5mm},
  boxed title style={tile,size=minimal,left=0.2mm,right=0.5mm,
    colback=white,before upper=\strut},
  title=#2,#1
}

\newcommand{\tool}{\textsc{SpecMind}\xspace}

\definecolor{darkred}{rgb}{0.6,0,0}

\newcommand{\code}[1]{{\footnotesize\texttt{#1}}}

\title{{\tool}: Cognitively Inspired, Interactive Multi-Turn Framework for Postcondition Inference}

\author{
 \textbf{Cuong Chi Le\textsuperscript{1}},
 \textbf{Minh V.T. Pham\textsuperscript{2}},
 \textbf{Tung Duy Vu\textsuperscript{3}},
 \textbf{Cuong Duc Van\textsuperscript{2}},
\\
 \textbf{Huy N. Phan\textsuperscript{2}},
 \textbf{Hoang N. Phan\textsuperscript{4}},
 \textbf{Tien N. Nguyen\textsuperscript{1}}
\\
\\
 \textsuperscript{1}The University of Texas at Dallas,
 \textsuperscript{2}FPT Software AI Center,
 \\
 \textsuperscript{3}VinUniversity,
 \textsuperscript{4}Nanyang Technological University
\\
 \small{
   \textbf{Contacts:} {\{cuong.le, tien.n.nguyen\}@utdallas.edu}
 }
}

\begin{document}
\maketitle
\begin{abstract}
Specifications are vital for ensuring program correctness, yet writing them manually remains challenging and time-intensive. 
Recent large language model (LLM)-based methods have shown successes in generating specifications such as postconditions, but existing single-pass prompting often yields inaccurate results. In this paper, we present {\tool}, a novel framework for postcondition generation that treats LLMs as interactive and exploratory reasoners rather than one-shot generators. {\tool} employs feedback-driven multi-turn prompting approaches, enabling the model to iteratively refine candidate postconditions by incorporating implicit and explicit correctness feedback, while autonomously deciding when to stop. This process fosters deeper code comprehension and improves alignment with true program behavior via exploratory attempts. Our empirical evaluation shows that {\tool} significantly outperforms state-of-the-art approaches in both accuracy and completeness of generated postconditions.
\end{abstract}

\section{Introduction}
\label{sec:intro}

Program specifications (pre-/postconditions) are central to checking that program behavior matches intent, but writing them manually is tedious. Prior work on automated inference broadly falls into four categories. First, program-analysis methods infer invariants dynamically from executions~\cite{BeschastnikhBSSE2011,daikon99}—and thus are coverage-limited—or statically from code, often at the cost of conservative, false-positive–prone results~\cite{ramanathan-pldi07,yiwei-icse11}. Second, data-mining approaches extract common API-usage patterns (e.g., call pairs/sequences and automata) from large codebases~\cite{zeller07,taoxie-ase09,mike-ase09,zeller-ase09,mapo09}, but typically do not infer semantic specifications such as pre-/postconditions.

Third,  EvoSpex~\cite{10.1109/ICSE43902.2021.00112} uses execution-guided evolutionary search to evolve postcondition candidates, yet its handcrafted operators 
only weakly exploit program semantics, leading to incomplete or brittle postconditions. Fourth, \code{nl2postcond}~\cite{nl2postcond} leverages LLMs to translate code and informal documentation into postconditions. The limitation of simple prompting in \code{nl2postcond} lies in their reliance on {\em single-pass generation}: \code{nl2postcond} expects the LLM to generate a correct and complete postcondition in a single prompt.
Such prompting results in {\em incorrect postconditions} that are syntactically plausible but semantically inaccurate, failing correctness tests or lacking discriminative power for bug detection.
(Related work details are given in Appendix~\ref{sec:related}).

\begin{figure*}[t]
    \centering
    \includegraphics[width=0.8\linewidth]{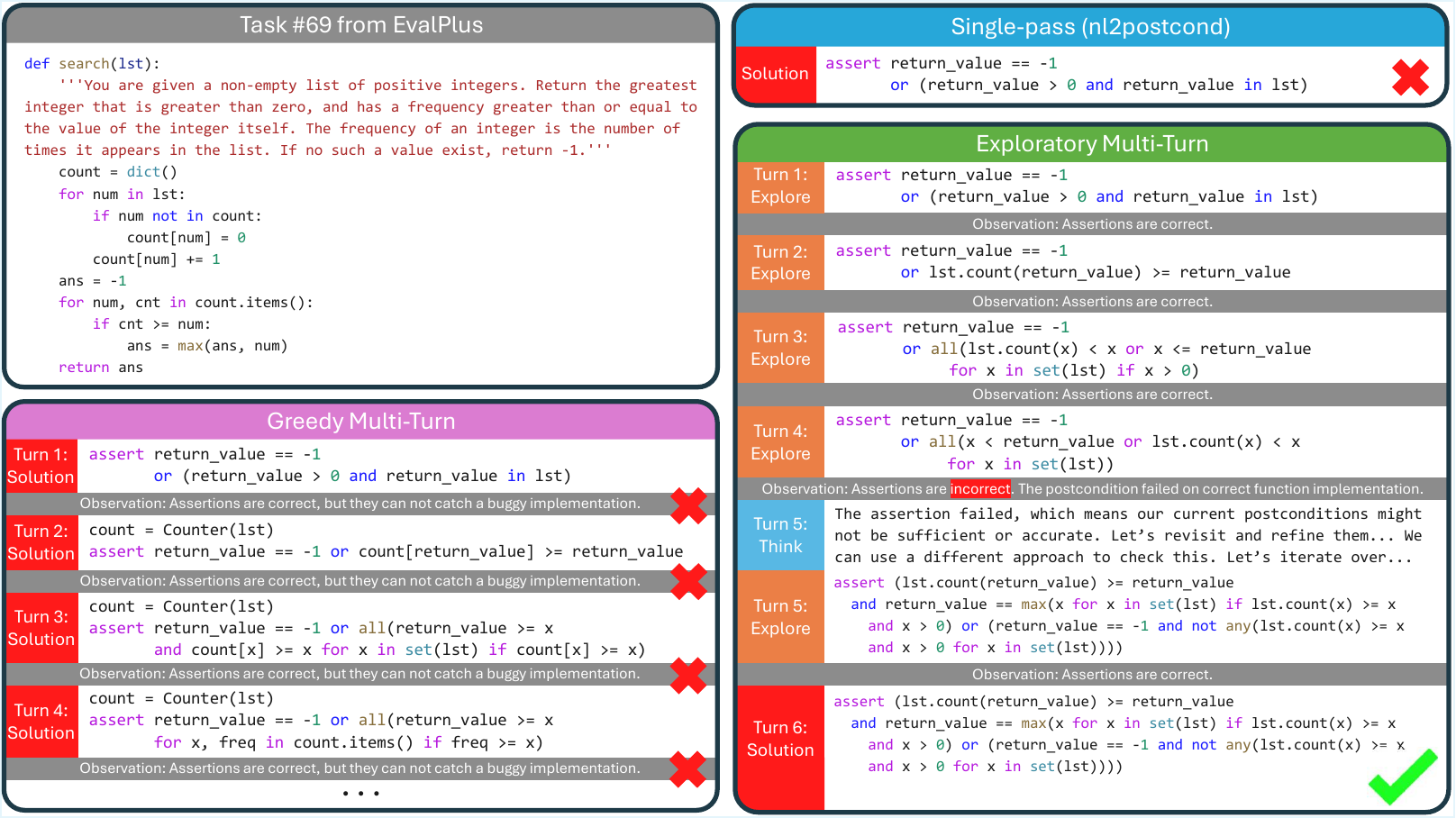} 
    \vspace{-3pt}
    \caption{
Example of task \#69 from EvalPlus with postconditions from Single-pass (nl2postcond~\cite{nl2postcond}), Greedy Multi-turn, and Exploratory Multi-turn. Blue blocks show model reasoning (omitted for all but turn 5 of Exploratory Multi-turn for space reason), red, orange, and gray blocks show submission and exploration attempts, and feedback. {\textcolor{green}{\checkmark}}: correct and complete postcondition, {\textcolor{red}{$\times$}}: otherwise.
}
    \label{fig:motiv1}
\end{figure*}

In our work, {\tool}, we aim to enhance the LLM's capability in generating postconditions from  given source code. A more robust alternative is {\em retry-based prompting} ({\bf Greedy Multi-turn} or Greedy for short), where the LLM is iteratively asked to regenerate postconditions until one passes a soundness check and exceeds a certain threshold of completeness. While this approach introduces feedback into the loop, it remains {\em reactive and externally controlled} -- the LLM is not actively involved in the reasoning process but simply retries in response to failure signals.
%
Thus, we also propose {\bf Exploratory Multi-Turn} (Exploratory for~short), offering a more cognitively aligned framework in which the LLM is treated not just as a {\em code-to-code generator}, but as an {\bf interactive reasoner}. We prompt the LLM to autonomously perform the exploration, internalize the reasoning process to derive the postcondition, and decide by itself when to stop the process. {\em We prompt the model~to perform one or multiple turns for exploration}, i.e.,~submitting postconditions to obtain the intermediate feedback on soundness and completeness, before deciding to submit the final postcondition. {\em This method engages the model in an iterative self-refinement process where it internalizes the feedback, reflects on prior attempts with the observations on the results of the intermediate turns, and gradually converges toward a more accurate and comprehensive postcondition}. This feedback to LLMs can be implicit (e.g., soundness and/or completeness signals) or explicit (e.g., un-caught mutants). Greedy is a special case of Exploratory Multi-turn where no intermediate turn of exploration is allowed and the model submits the final result after each attempt. {\color{custom-blue}{While Greedy is an iterative algorithm enhanced with history, Exploratory Multi-turn is an {\em algorithmically defined, iterative, exploratory reasoning framework} with 1) explicit exploration-submission decoupling, 2) structured history-aware reasoning, 3) best-so-far tracking, and 4) autonomous stopping}} (Section~\ref{sec:motivation}).
Those characteristics give Exploratory Multi-turn the following key benefits:

{\em Deeper Code Comprehension}: By interacting with feedback and refining its outputs over multiple turns, the model is encouraged to form a more semantically grounded understanding of the code logic, rather than relying on superficial patterns.


{\em Autonomous Reasoning}: In the autonomous exploration, the model determines when to stop, which introduces a degree of self-awareness and decision-making on the sufficiency of its~answer.



We evaluated our model on EvalPlus \cite{nl2postcond} and FixEval \cite{fixeval}. Exploratory Multi-turn achieves 99.4\% correctness and 89.6\% completeness, relatively improving over the baseline \code{nl2postcond} by 26.1\% in correctness and 2.48X in completeness. The average increase in completeness after one submission attempt of Exploratory Multi-Turn is 1.67X higher than that of Greedy. {\em The postconditions from {\tool} can detect 1.42X-2.14X more bugs than \code{nl2postcond} in FixEval with less costs}. Our data and code can be found at~\cite{Specmind}.






\section{Illustration}
\label{sec:motivation}

The function \code{search} in Fig.~\ref{fig:motiv1} {\em takes a non-empty list of positive integers as input and returns the greatest integer in the list whose frequency is greater than or equal to its value; if no such integer exists, it returns $-1$}. In the ground truth, the complete specification for this function must contain all of the following sub-conditions:

(1) the return value is either $-1$ or a positive integer from the input list; 

(2) if the return value is positive, its frequency must be greater than or equal to its value; 

(3) if the return value is positive, it must be the greatest integer satisfying the frequency condition;

(4) if the return value is $-1$, no integer in the list satisfies the frequency condition. 


We applied the state-of-the-art, LLM-based, postcondition generation tool, \code{nl2postcond}~\cite{nl2postcond} on the code of the \code{search} function. 
\code{nl2postcond} leverages a {\em single-pass} prompting strategy to request the LLM used in this example, Llama 4 Scout, to generate the postcondition.


A candidate postcondition can be evaluated using two metrics: {\em correctness and completeness}~\cite{nl2postcond}. {\bf Correctness} refers to if the postcondition passes all test cases in the given test set. A postcondition is correct if it holds true for all test cases that the original function passes (no contradiction to the behavior in test cases). However, correctness alone is not sufficient as a trivial postcondition like \code{"True"} is always correct but useless in bug detection. {\bf Completeness}, referring to discriminative power, measures how informative the postcondition is. It reflects the postcondition’s ability to detect incorrect versions of the function (e.g., mutated code). This is done by checking if the postcondition can "catch" mutants (i.e., detecting bugs), i.e., the~modified versions that behave~differently. 
Formally, completeness is the percentage of mutants that can be distinguished (i.e., rejected) by the postcondition when used with the test suite. 

From our experiment on the above example, we made the following observations:

1. The state-of-the-art LLM-based, {\bf single-pass} approach, \code{nl2postcond}~\cite{nl2postcond},
generated the postconditions ensuring that the return value is either -1 or a positive integer from the input list, thereby satisfying the condition (1). However, none of the conditions (2)--(4) was generated.



2. {\bf Feedback-driven Multi-Turn Approaches}. Prior work suggests that the feedback helps LLMs ground their generations more effectively~\cite{gehring2025rlef}. Building on this, we integrated feedback into the process by explicitly indicating which generated constraints were incorrect or non-discriminative. Thus, we propose the following two feedback-driven Multi-Turn approaches: 

{\bf Greedy Multi-Turn.} The first one is {\color{custom-blue}{\em a feedback-guided, test-driven iterative prompting for refinement}}. We employed a simple feedback mechanism in which the LLM {\em repeatedly attempted to produce a correct postcondition} until all test cases passed and a desired completeness is reached, using feedback from previous failed attempts. This feedback includes either binary pass/fail signals or syntax errors. The iterative process enabled the model to progressively cover the condition (1) in the first attempt, the condition (2) in the second, and the condition (3) in the third, eventually combining them into a single assertion by the fourth attempt. Within the 12-attempt limit, the model focused predominantly on these three conditions, but failed to generate the condition (4). The generated assertions consistently began with \code{assert return\_value == -1} or ..., a pattern that can be trivially satisfied by faulty code returning -1 in all~cases.

{\bf Exploratory Multi-turn.} In real-world learning, students rarely solve complex problems correctly in a single attempt. Effective learners instead ask clarifying questions, explore partial solutions, and test their understanding before committing to a final answer. This differs from blindly retrying after a “try again” signal, and from single-pass solving without refinement. Inspired by this, we propose Exploratory Multi-turn for postcondition generation, where the LLM acts not as a code-to-code generator but as an \textbf{active reasoner}. It autonomously explores intermediate reasoning paths, probes partial answers, internalizes feedback, and decides when it has enough evidence to produce the result.




In the example, similar to Greedy Multi-turn, the first three attempts correctly generated conditions (1)–(3). In the 4th attempt, a minor change triggered an \code{AssertionError}. Unlike Greedy, which often stagnates on the same incomplete assertion, Exploratory Multi-Turn broke from its prior pattern and shifted direction in the fifth attempt. The model explicitly noted: \textit{“The assertion failed, which means our current postconditions might not be sufficient or accurate. Let’s revisit and refine them... We can use a different approach... Let’s iterate over...”}. This shift produced a postcondition that covered condition (4) and discarded the overly permissive pattern \code{assert} \code{return\_value} == -1 or .... It finally generated the correct and complete postcondition, suggesting that Exploratory Multi-Turn helps escape local optima and improves completeness.

{\color{custom-blue}{
Specifically, our key design ideas are as follows:

1. {\em Exploration-submission decoupling}: we explicitly decouple exploration (hypothesis probing via <assert>) from commitment (<solution>). Only submission attempts update the best-so-far candidate, while exploratory turns gather semantic evidence. This separation  transforms the process from feedback-guided heuristic search into structured exploring with hypothesis probing. 

2. {\em Best-so-far tracking:} {\tool} uses a completeness threshold, best-so-far tracking, and a model-controlled stop decision that can terminate when the threshold is met, returns diminish, or the attempt budget is reached. This ensures the final output is the strongest candidate observed (best-so-far), rather than simply the last attempt.

3. {\em History-aware reasoning}: 
In iterative prompting, each attempt is often independent or loosely conditioned on prior error messages. To support exploration, {\tool} maintains a structured history buffer that is embedded into the prompt. This enables explicit reflection on prior mistakes and behavioral gaps during historical exploration, rather than relying solely on the immediate feedback. 

4. {\em Autonomous stopping mechanism}: {\tool} allows the model to determine termination via LLM, based on improvement trajectory and completeness threshold. This introduces a self-regulated search dynamic absent in retry-based iterative prompting paradigms.

}}

\section{Problem Formulation}
\label{sec:formulation}


Given a function $f$:$\mathcal{I} \rightarrow \mathcal{O}$, where $\mathcal{I}$ and $\mathcal{O}$ are the input  and output domains, the goal is to synthesize a postcondition $\phi: \mathcal{I} \times \mathcal{O} \rightarrow \{\code{true}, \code{false}\}$ that characterizes the correct behavior of $f$.



The postcondition $\phi$ is a logical predicate that must hold for any input-output pair $(i, o)$ where $o = f(i)$. Formally: $\forall i \in \mathcal{I}, \quad \phi(i, f(i)) = \texttt{true}.$
In practice, $\phi$ is often expressed as an assertion or a condition over program variables after execution. The goal is to infer such a $\phi$ automatically, using the following inputs: a function’s source code, a finite test suite $\mathcal{T} = \{(i_1, o_1), \dots, (i_n, o_n)\}$, and a set of mutants $\mathcal{M}$ of the original $f$. Two evaluation criteria for a synthesized postcondition $\phi$ include



{\bf Correctness (Corr.)}: The postcondition must hold on all known correct executions captured in the test suite: $\forall (i_j, o_j) \in \mathcal{T}, \quad \phi(i_j, o_j) = \code{true}.$ The postcondition must reject incorrect variants.

\vspace{2pt}
{\bf Completeness (Comp.)}: Given a set of program mutants $\mathcal{M} = \{f'_1, f'_2, \dots, f'_m\}$, the completeness of a postcondition measures the proportion of mutants for which the postcondition fails:
{\small
$$
\emph{Comp}(\phi) =
\frac{\left| \{ f'_k \in \mathcal{M} \mid \exists i \in \mathcal{I},\ \phi(i, f'_k(i)) = \code{false} \} \right|}{|\mathcal{M}|}
$$
}

Correctness ensures consistency with the given code, while completeness reflects the specification's robustness in identifying deviations/defects.

\vspace{2pt}
\noindent {\bf Objective of Postcondition Inference/Generation.}
Given a function $f$, a test suite $\mathcal{T}$, and a set of mutants $\mathcal{M}$, the objective of {\tool} is to synthesize a postcondition $\phi$ that:

1. Achieves full correctness over $\mathcal{T}$;

2. Maximizes completeness over $\mathcal{M}$;



\section{Exploratory Multi-turn Algorithm}
\label{sec:algo}






\paragraph{Algorithm supporting information.}

{\em Completeness Score $s \in [0, 1]$}: Each feedback evaluation returns a quantitative measure of completeness (e.g., how much of the expected behavior is covered).

{\em Threshold $\tau \in [0, 1]$}: A predefined target completeness level. The loop continues until the model achieves this threshold or chooses to stop early.

{\em Best-so-far Tracking}: The algorithm keeps track of the best postcondition submitted by the model so far (based on the highest completeness score) and returns it if the loop ends prematurely.

\textit{LLM.decidesToStop.} At each attempt, the model may choose to submit a postcondition once it is confident after a process of exploration and refinement or reaches a maximum number of attempts. The submitted postcondition is then evaluated and, if it improves upon the \textit{Best-so-far Tracking} and satisfies the completeness threshold, the algorithm triggers \textit{LLM.decidesToStop} to terminate and return the best postcondition generated so far.

\begin{figure}
\centering
\scalebox{0.95}{
\begin{minipage}{3.2in}
\footnotesize
\begin{algorithmic}[1]
\Require Function code $f$, test suite $T$, completeness threshold~$\tau$
\Ensure Final postcondition $\phi_{\text{final}}$

\State $\mathcal{H} \gets [\,]$ \Comment{History buffer}
\State $\phi_{\text{best}} \gets \emptyset$, $s_{\text{best}} \gets 0.0$

\While{True}
    \State $prompt \gets \text{constructPrompt}(f, \mathcal{H})$
    \State $(\phi, type) \gets \text{LLM.generate}(prompt)$
    \Comment{$type \in \{\text{explore}, \text{submit}\}$}
    \State Append $(\phi, type)$ to $\mathcal{H}$ \Comment{Update history buffer}

    \If{$type = \text{submit}$} \Comment{$score \in [0,1]$}
        \State $(feedback, score) \gets \text{evaluateFeedback}(\phi, T)$ 
        \State \text{annotateLast}$(\mathcal{H}, feedback, score)$ \Comment{Feedback}

        \If{$score > s_{\text{best}}$}
            \State $\phi_{\text{best}} \gets \phi$, $s_{\text{best}} \gets score$ \Comment{Update best solution only on submission attempts}
        \EndIf
    \EndIf

    \If{$\text{LLM.decidesToStop}(\mathcal{H}, s_{\text{best}})$}
        \State \Return $\phi_{\text{best}}$ \Comment{When the termination condition is met; return best-so-far}
    \EndIf
\EndWhile
\end{algorithmic}

\vspace{-3pt}
\caption{Feedback-Driven Exploratory Multi-Turn Algorithm with Completeness Threshold}
\label{algo}
\end{minipage}}
\end{figure}

{\bf Details.} 
Fig.~\ref{algo} outlines Exploratory Multi-turn for a given function. We treat inference as an iterative process guided by structured feedback: {\tool} lets the LLM refine candidates until the completeness criterion is met or the submission budget is exhausted. We maintain a history buffer $\mathcal{H}$ that stores all attempted postconditions (exploratory and submitted) with metadata (line~1), and a best-so-far record $(\phi_{best}, s_{best})$ tracking the strongest submitted candidate (line~2).

Each iteration synthesizes a prompt from the function code $f$ and prior feedback in $\mathcal{H}$ (line~4), encouraging the model to reflect on earlier mistakes. It then produces either an exploratory refinement or a submission. Submitted candidates are evaluated by a feedback engine that returns qualitative outcomes (e.g., pass/fail) and a normalized completeness score $s\in[0,1]$ (line~6). We log the submission, feedback, and score in $\mathcal{H}$ (line~9), and update $(\phi_{best}, s_{best})$ if $s>s_{best}$ (lines~10–11). Termination is controlled by \code{LLM.decidesToStop} (line~12), which stops when a candidate reaches threshold $\tau$, returns diminish, or the maximum attempts $\mu$ is reached; the algorithm returns $\phi_{best}$ (line~13).

Overall, we use the LLM as an interactive reasoner that incorporates feedback and self-regulates the search, reducing hallucination and overgeneralization. Greedy Multi-turn (not shown) repeatedly prompts the LLM to resubmit until all tests pass or until $\tau$ or the attempt budget is reached.

\section{Prompt Design and Feedback}
\label{sec:prompt}

\begin{figure}
\centering
\scriptsize 
\begin{tcolorbox}[colback=gray!5,
                 colframe=black!70!white,
                 title={\textbf{Prompt Template}},
                 width=\linewidth,
                 boxsep=1pt, left=2pt, right=2pt, top=2pt, bottom=2pt]

\textbf{Objective:} Verify correctness of a Python function using its natural language description and implementation.  
Your goal is to write symbolic postconditions --- Python \texttt{assert} statements that validate specific behavioral properties of the function's return value.

\textbf{Turn Structure:}  
Each turn must begin with a \texttt{<think>} block containing reasoning about purpose, constraints, and edge cases.  
After this reasoning, choose exactly one of the following actions:

\begin{itemize}[itemsep=0pt,parsep=0pt,topsep=1pt,left=0.5em]
    \item \colorbox{green!40}{\texttt{<assert>} – Propose a single candidate assertion for testing.} \colorbox{green!40}{$\rightarrow$ Triggers an \texttt{<observation>} with feedback.}
    \item \texttt{<solution>} – Provide the final refined postcondition.
\end{itemize}
\textbf{You will now be given the function} \texttt{\{function\_name\}}:
\vspace{-0.5em}
\begin{lstlisting}[language=Python,numbers=none,basicstyle=\ttfamily\scriptsize]
{function_signature}
    {function_docstring}
    {function_implementation}
\end{lstlisting}
\vspace{-0.5em}
Let's begin
\vspace{-0.5em}
\begin{tcolorbox}[colback=blue!2!white,
                 colframe=black!30!white,
                 coltext=black,
                 title={\textbf{Turn 1: $(\phi_1, \text{feedback}_1)$}},
                 left=1pt,right=1pt,top=1pt,bottom=1pt,boxsep=1pt]
\tiny
\textcolor{blue}{\texttt{<think>} ... \texttt{</think>}}  
\textcolor{orange}{\texttt{<assert>} $\phi_1$ \texttt{</assert>}}  
\textcolor{yellow!60!black}{\texttt{<observation>} $\text{feedback}_1$ \texttt{</observation>}}
\end{tcolorbox}

\texttt{...}

\begin{tcolorbox}[colback=orange!2!white,
                 colframe=black!30!white,
                 coltext=black,
                 title={\textbf{Turn $t{-}1$: $(\phi_{t-1}, \text{feedback}_{t-1})$}},
                 left=1pt,right=1pt,top=1pt,bottom=1pt,boxsep=1pt]
\tiny
\textcolor{blue}{\texttt{<think>} ... \texttt{</think>}}  
\textcolor{orange}{\texttt{<assert>} $\phi_{t-1}$ \texttt{</assert>}}  
\textcolor{yellow!60!black}{\texttt{<observation>} $\text{feedback}_{t-1}$ \texttt{</observation>}}
\end{tcolorbox}
\end{tcolorbox}
\vspace{-6pt}
\caption{
Template prompt for Greedy and Exploratory.}
\scriptsize{\textit{\colorbox{green!40}{The highlighted \texttt{<assert>} action is specific to  \textbf{Exploratory Multiturn}.}}}
\label{fig:prompt-template}
\end{figure}


\textbf{Prompt Design}. We adopt a modular and evolving prompt template that reflects the model’s current understanding, errors in prior attempts, and the functional structure of the code under analysis. The base prompt includes the full code of the target function and, optionally, a high-level natural language description (if available). Moreover, each subsequent prompt incorporates an interaction history consisting of prior postconditions, associated feedback, and explicit instructions for refinement. Formally, let the function be denoted $f$, and let the interaction history up to turn $t$ be represented~as:
\begin{equation}
\label{eq1}
\begin{aligned}
\mathcal{H}_t =\;&
[( \phi_1, \text{feedback}_1 ), ( \phi_2, \text{feedback}_2 ), \dots, \\
&\quad ( \phi_{t-1}, \text{feedback}_{t-1} )] \nonumber
\end{aligned}
\end{equation}
The prompt at turn $t$, denoted $\text{Prompt}_t$, is synthesized as:
$
\text{Prompt}_t = \code{PromptTemplate}(f, \mathcal{H}_t)
$
where \code{PromptTemplate} is a function that formats code, prior generations, and critiques into an LLM-readable input. A prompt template is shown in Fig.~\ref{fig:prompt-template}.

Our evolving structure allows the LLM to internalize past mistakes and progressively refine its output. Unlike static prompt, our dynamic history-aware formulation fosters deeper reasoning and removes the need to reset the context between turns.

\textbf{Feedback Mechanism}. Our feedback mechanism serves two goals: 1) guiding the LLM’s refinement with actionable signals, and 2) providing a quantitative estimate of each candidate’s semantic completeness. Each LLM-generated postcondition $\phi_t$ is evaluated against the code’s observed behavior, typically using a test suite and a set of~mutants.

The feedback $\code{feedback}_t$ includes: (a) a \textbf{correctness indicator}--whether $\phi_t$ holds on all tests; and (b) a \textbf{completeness score}--the fraction of mutants “caught” by $\phi_t$ (i.e., faulty variants that violate the postcondition). If the completeness threshold is~not reached within the attempt budget, we also return the remaining uncaught mutants as additional feedback. 
We append the prompt–feedback pair to the history $\mathcal{H}_t$ to steer subsequent generations toward remaining behavioral gaps.
If available, we can use symbolic execution or verification backends; otherwise, correctness is checked empirically on tests and completeness is measured via mutant catching. 
Grounding prompts in correctness and completeness yields an informative refinement loop.

\section{Empirical Evaluation}
\label{sec:eval}

\begin{table*}[t]
\vspace{-6pt}
\scriptsize
\centering
\caption{
Postcondition Generation Effectiveness with Llama 4 (RQ1). $\tau$: completeness threshold, $\mu$: max turns; R. Sampl.: run nl2postcond $\mu$ independent times, Subs: avg submissions, Corr: correctness, Comp.: Completeness.
}
\vspace{-6pt}
\tabcolsep 1.7pt
\label{tab:rq1}
\begin{minipage}{0.49\textwidth}

\begin{tabular}{l|c|c|c|c|c|}
\toprule
\textbf{Method} & \textbf{Config.} & \textbf{Attempts} & \textbf{Avg Subs} & \textbf{Corr.} & \textbf{Comp.} \\
 & & \textbf{min-max} & \textbf{min-max} &  &  \\
\midrule
\multirow{3}{*}{%
  \begin{tabular}{@{}c@{}}
  nl2postcond \\
  (Baseline)
  \end{tabular}
} 
  & \multirow{3}{*}{Single-pass} 
  & \multirow{3}{*}{1.0} 
  & \multirow{3}{*}{1.0} 
  & \multirow{3}{*}{73.3\%} 
  & \multirow{3}{*}{36.0\%} \\
& & & & & \\
& & & & & \\
\hline
\hline
R.Sampl. w. nl2postcond & \multirow{3}{*}{%
  \begin{tabular}{@{}c@{}}
  $\tau$ = 50 \\
  $\mu$ = 4
  \end{tabular}
} & 3.0 (1 - 4) & 3.0 (1 - 4) & 79.2\% & 41.2\% \\
Greedy & & 2.3 (1 - 4) & 2.3 (1 - 4) & 93.7\% & 69.2\% \\
Exploratory & & 3.8 (3 - 4) & 1.0 (1 - 2) & 84.3\% & 70.6\% \\
\hline
R.Sampl. w. nl2postcond & \multirow{3}{*}{%
  \begin{tabular}{@{}c@{}}
  $\tau$ = 50 \\
  $\mu$ = 8
  \end{tabular}
} & 5.1 (1 - 8) & 5.1 (1 - 8) & 84.9\% & 47.9\% \\
Greedy & & 2.9 (1 - 8) & 2.9 (1 - 8) & 96.9\% & 70.7\% \\
Exploratory & & 5.7 (3 - 8) & 1.2 (1 - 3) & 93.1\% & 81.5\% \\
\hline
R.Sampl. w. nl2postcond& \multirow{3}{*}{%
  \begin{tabular}{@{}c@{}}
  $\tau$ = 50 \\
  $\mu$ = 12
  \end{tabular}
} & 6.9 (1 - 12) & 6.9 (1 - 12) & 84.9\% & 50.9\% \\
Greedy & & 3.3 (1 - 12) & 3.3 (1 - 12) & 100.0\% & 75.0\% \\
Exploratory & & 6.5 (3 - 12) & 1.3 (1 - 5) & 98.1\% & 86.4\% \\
\hline
\hline
R.Sampl. w. nl2postcond & \multirow{3}{*}{%
  \begin{tabular}{@{}c@{}}
  $\tau$ = 70 \\
  $\mu$ = 4
  \end{tabular}
} & 3.5 (1 - 4) & 3.5 (1 - 4) & 79.2\% & 41.3\% \\
Greedy & & 2.6 (1 - 4) & 2.6 (1 - 4) & 95.6\% & 68.6\% \\
Exploratory & & 3.8 (3 - 4) & 1.0 (1 - 2) & 84.3\% & 70.4\% \\
\hline
\end{tabular}

\end{minipage}
\hfill
\begin{minipage}{0.49\textwidth}

\begin{tabular}{|l|c|c|c|c|c|}
\toprule
\textbf{Method} & \textbf{Config.} & \textbf{Attempts} & \textbf{Avg Subs} & \textbf{Corr.} & \textbf{Comp.} \\
 & & \textbf{min-max} & \textbf{min-max} & & \\
\midrule
R.Sampl. w. nl2postcond& \multirow{3}{*}{%
  \begin{tabular}{@{}c@{}}
  $\tau$ = 70 \\
  $\mu$ = 8
  \end{tabular}
} & 6.4 (1 - 8) & 6.4 (1 - 8)  & 84.9\% & 48.4\% \\
Greedy & & 3.8 (1 - 8) & 3.8 (1 - 8) & 96.9\% & 78.6\% \\
Exploratory & & 5.7 (2 - 8) & 1.2 (1 - 4) & 93.1\% & 83.7\% \\
\hline
R.Sampl. w. nl2postcond & \multirow{3}{*}{%
  \begin{tabular}{@{}c@{}}
  $\tau$ = 70 \\
  $\mu$ = 12
  \end{tabular}
} & 9.0 (1 - 12) & 9.0 (1 - 12) & 84.9\% & 51.9\% \\
Greedy & & 5.3 (1 - 12) & 5.3 (1 - 12) & 100.0\% & 78.4\% \\
Exploratory & & 6.5 (3 - 12) & 1.4 (1 - 6) & 97.5\% & 85.9\% \\
\hline
\hline
R.Sampl. w. nl2postcond & \multirow{3}{*}{%
  \begin{tabular}{@{}c@{}}
  $\tau$ = 90 \\
  $\mu$ = 4
  \end{tabular}
} & 3.7 (1 - 4) & 3.7 (1 - 4) & 79.2\% & 41.5\% \\
Greedy & & 3.2 (1 - 4) & 3.2 (1 - 4) & 96.9\% & 76.4\% \\
Exploratory & & 3.8 (3 - 4) & 1.1 (1 - 2) & 84.3\% & 71.6\% \\
\hline
R.Sampl. w. nl2postcond & \multirow{3}{*}{%
  \begin{tabular}{@{}c@{}}
  $\tau$ = 90 \\
  $\mu$ = 8
  \end{tabular}
} & 7.1 (1 - 8) & 7.1 (1 - 8) & 84.9\% & 48.9\% \\
Greedy & & 5.1 (1 - 8) & 5.1 (1 - 8) & 99.4\% & 81.1\% \\
Exploratory & & 6.2 (3 - 8) & 1.4 (1 - 4) & 94.3\% & 83.2\% \\
\hline
R.Sampl. w. nl2postcond & \multirow{3}{*}{%
  \begin{tabular}{@{}c@{}}
  $\tau$ = 90 \\
  $\mu$ = 12
  \end{tabular}
} & 10.3 (1 - 12) & 10.3 (1 - 12) & 84.9\% & 52.6\% \\
Greedy & & 6.2 (1 - 12) & 6.2 (1 - 12) & 98.7\% & 85.8\% \\
Exploratory & & 7.2 (3 - 12) & 1.7 (1 - 6) & {\bf 99.4\%} & {\bf 89.6\%} \\
\bottomrule
\end{tabular}
\end{minipage}
\end{table*}

We aim to answer the following questions:

\textbf{RQ1.} [Effectiveness--Efficiency]. How effecti\-ve and efficient is our tool in postcondition~inference?

\textbf{RQ2.} [Reasoning]. How does {\tool} with Exploratory Multi-turn perform reasoning?


\textbf{RQ3.} [Stratifying Results]. How does {\tool} with feedback perform on hard cases?



{\bf RQ4.} [Cost Efficiency]. How efficient is {\tool} in token costs and actual costs?

{\color{custom-blue}{{\bf RQ5.} [Bug Detection]. How effective is {\tool} in bug detection?}}




\vspace{1pt}
\noindent \textbf{Benchmarks}. We selected EvalPlus as our dataset as it was used in \code{nl2postcond}~\cite{nl2postcond}. It has 164 Python problems, each with a function stub, textual description, reference implementation, and validation tests \cite{evalplus}. EvalPlus updates the HumanEval benchmark \cite{chen2021evaluating}, containing the same problems but with more extensive test suites ($\approx$ 775 test cases per problem). 


\noindent \textbf{Baselines}. We chose as baseline the state-of-the-art \code{nl2postcond}, with access to the reference implementation. \code{nl2postcond} was shown to outperform prior approaches such as TOGA~\cite{dinella2022toga} and Daikon~\cite{daikon99}. We used Llama~4~Scout as the underlying LLMs for the tools in all experiments.


\vspace{2pt}
\noindent \textbf{Evaluation Metrics}. We use two key metrics: 1) {\em Correctness} refers to whether the postcondition passes all test cases in the given test suite; 2) {\em Completeness} is the percentage of mutants distinguishable (i.e., rejected) by the postcondition given the test suite. {\color{black}{We used the same mutant set from \code{nl2postcond} for a fair comparison.}}

To calculate the efficiency over submission attempts, we define the efficiency score in a setting as $\mathbf{E} = \frac{1}{N} \sum_{i=1}^{N} \frac{\text{completeness score}_i}{\text{number of submission turns}_i}$, where $N$ is the number of functions in the dataset. The efficiency score $\mathbf{E}$ measures the average increase of completeness after one submission attempt.


\vspace{-5pt}
\subsection{Effectiveness and Efficiency (RQ1)}
As seen in Table~\ref{tab:rq1}, with Exploratory achieving peak performance at $\tau$=90, $\mu$=12 ({\bf 99.4\% correctness, 89.6\% completeness}) while requiring an average of 7.2 attempts and 1.7 submissions per task. \code{nl2postcond} achieves 73.3\% correctness and 36.0\% completeness using a single attempt.

Across most settings, for both Greedy and Exploratory Multi-turn, despite allowing a higher maximum number of attempts, {\em the average number of attempts range between 2.3-7.2, with the correctness increasing from 84.3\% to 99.4\% and completeness increasing from 69.2\% to 89.6\%}. The higher maximum number of attempts allowed, the higher correctness and completeness for Greedy and Exploratory Multi-turn. For Exploratory Multi-turn, among 3.8-7.2 attempts, it explores 3.5 times on average and submits the resulting postcondition from 1.0 to 1.7 times. As seen, on average, {\tool} uses {\em a small number of attempts} and does not need to use up the attempt budget. However, the highest numbers of attempts reach the maximum budget. That means that {\em it needs to use all the budget of attempts in hard cases} (see RQ3). The results run on GLM 4.7 is shown in Appendix~\ref{sec:appendix_glm}.

\subsubsection {\bf Correctness}


We first study the effect of the attempt budget $\mu$. 
With $\mu$=$4$, both Exploratory and Greedy Multi-turn outperform \code{nl2postcond}. Under the most relaxed setting ($\tau$=50, $\mu$=4), Greedy reaches 93.7\% correctness in 2.3 attempts on average, a 20-point gain over \code{nl2postcond} (73.3\%). Exploratory also improves over \code{nl2postcond}, but achieves 84.3\% due to limited exploratory turns.
Exploratory benefits more from larger $\mu$: the average number of functions with correct postconditions rises from 139.0 ($\mu=4$) to 153.7 ($\mu$=8) and 161.3 ($\mu$=12). Greedy improves less with $\mu$. On average, increasing $\mu$ by 4 yields a 6.8\% correctness gain for Exploratory versus 2.1\% for Greedy, as Exploratory can probe more candidates before submission (Section~\ref{sec:rq2}).

For comparison, we also evaluate \textbf{Random Sampling} with \code{nl2postcond}, i.e., querying the LLM independently $\mu$ times with early stopping when $\tau$ is met. Random Sampling achieves 79.2\% correctness at ($\tau$=50, $\mu$=4) and peaks at 84.9\% under the strictest setting ($\tau$=$90$, $\mu$=$12$). Under the same setting, Greedy and Exploratory reach 98.7\% and 99.4\% correctness, using 6.2 and 7.2 attempts on average, compared to 10.3 for Random Sampling.

\subsubsection{\bf Completeness} Exploratory Multi-turn outperforms Greedy in all settings except $\tau{=}90,\mu{=}4$. The smallest~completeness gap is at $\tau{=}50,\mu{=}4$ (70.6\% vs. 69.2\%), and the largest at $\tau{=}50,\mu{=}12$ (86.4\% vs. 75.0\%). Our tool's lowest completeness is 68.6\% at $\tau{=}70,\mu{=}4$, far above single-pass \code{nl2postcond} (36.0\%), highlighting the benefit of multi-turn refinement for more discriminative specifications. Random Sampling improves with more attempts (e.g., at $\tau{=}90$, 41.5\% with $\mu{=}4$ to 52.6\% with $\mu{=}12$) but remains well below {\tool}’s worst case (68.6\%). {\color{black}{We also analyze completeness scores across submission attempts.
See Appendix~\ref{sec:appendix_patter} for details.


\begin{figure}[t]
\centering
    \includegraphics[width=0.37\textwidth]{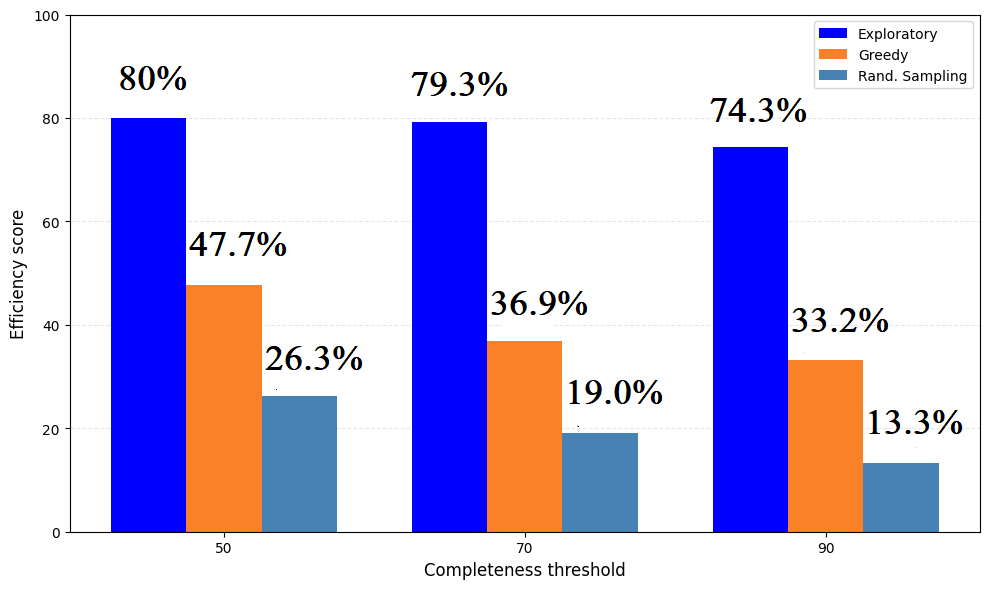}
    \vspace{-6pt}
    \caption{Efficiency for configurations with $\mu$=12.}
    \label{fig:efficiency}
\end{figure}

\begin{table*}[ht]
\centering
\small
\tabcolsep 3pt
\caption{Exploratory Multi-turn's reasoning behavior categories with descriptions, frequencies, examples (RQ2).}
\label{tab:reasoning_categories}
\vspace{-8pt}
\begin{tabular}{l@{\hskip 7pt}p{5.2cm}@{\hskip 6pt}c@{\hskip 4pt}p{5cm}}
\toprule
\textbf{Reasoning Category} & \textbf{Description} & \textbf{Freq} & \textbf{Example} \\
\midrule
\footnotesize {\em Return Value Type \& Range} & \footnotesize Checking the type and range of the return value. & \footnotesize 77 & \scriptsize "The assertion I proposed checks if … To further strengthen the assertion, I can check if the return value is a list of strings…" \\
\footnotesize {\em Check Base Cases} & \footnotesize Checking the base cases of a function. & \footnotesize 32 & \scriptsize "Let's check if the function returns the correct values for the base cases. For \( n=0 \), the return value should be 0…" \\
\footnotesize {\em Check Edge Cases} & \footnotesize Adding postconditions checking edge cases: empty inputs or boundary values. & \footnotesize 63 & \scriptsize "Let's consider the case where the input number is an integer. The decimal part should be exactly 0…" \\
\footnotesize {\em Postcondition Combination} & \footnotesize Combining multiple postconditions into one. & \footnotesize 59 & \scriptsize "Both properties seem to be correct… Let's combine these insights into a single assertion…" \\
\footnotesize {\em Refine Assertions} & \footnotesize Refining assertions to make them more concise, meaningful, or comprehensive. & \footnotesize 419 & \scriptsize "The combined assertion seems correct and comprehensive. However, let's make it more concise and readable…" \\
\footnotesize {\em Verify Function Behavior} & \footnotesize Adding assertions to verify correctness of calculations. & \footnotesize 270 & \scriptsize "The first assertion checks… Next, we want to check that the return value, when added to apples and blues, equals the total number of fruits…" \\
\footnotesize {\em Submit Final Solution} & \footnotesize Submitting assertions as the final solution without further refining. & \footnotesize 129 & \scriptsize "The assertion is now concise and readable. It correctly checks…, let's consider submitting it..." \\
\bottomrule
\end{tabular}
\label{tab:rq2}
\end{table*}

\subsubsection{\bf Efficiency} Fig.~\ref{fig:efficiency} shows that Exploratory Multi-turn is more submission-efficient than Greedy, peaking at 80.0\% vs. 47.7\%. That is, the average completeness gain per submission is \textbf{1.67$\times$} higher for Exploratory. Raising the completeness threshold increases final completeness but reduces efficiency: from $\tau{=}50$ to $\tau{=}90$, efficiency drops by 5.7\% for Exploratory and 14.5\% for Greedy, suggesting that more discriminative postconditions require more exploratories.
Compared to Random Sampling, Exploratory is more efficient (80.0\% vs. 26.3\%) and degrades less as $\tau$ increases. As $\tau$ rises from 50-70-90, Exploratory achieves 3.04$\times$, 4.17$\times$, and 5.59$\times$ the efficiency of Random Sampling. {\color{black}{See a case study comparing two approaches in Appendix~\ref{sec:appendix_case}. 
}}

\subsection{{\tool}'s Reasoning at Each Attempt}
\label{sec:rq2}



This experiment analyzes {\tool}’s reasoning behavior across attempts, focusing on Exploratory Multi-turn because each attempt may depend on prior exploratory or submission turns (Greedy Multi-turn contains only submission turns). We use $\tau{=}70$ since $\tau{=}50$ often stops too early and $\tau{=}90$ frequently exhausts the budget without meeting the threshold. We set $\mu{=}12$ to observe behavior over more attempts. We analyze all natural-language reasoning texts produced per attempt (enclosed by \code{<think>}…\code{</think>}), totaling 1,026 instances; each case has 6.5 reasoning instances on average.

To categorize reasoning behaviors, we first use an LLM to cluster reasoning texts in five batches (200–226 each), producing distinct, non-overlapping categories. We then manually consolidate and refine these into seven observable categories. Finally, we use an LLM to classify all reasoning texts into the seven categories for consistent labeling, and manually validate the results by inspecting a random 10\% sample per category.

Table \ref{tab:rq2} presents the categories of {\tool}’s reasoning along with their descriptions, frequencies, and representative examples. The most frequent behavior is \textit{Refine Assertions}, appearing in 419 reasonings, where the model improves existing assertions to make them more concise and comprehensive. Typical phrases include: {\em “To refine our postconditions...”}, {\em “A more accurate property might be...”}, or {\em “The current postcondition does not accurately reflect...”}. The second most frequent category is \textit{Verify Function Behavior}, with 270 reasonings. In this category, the model proposes assertions to verify functional properties. 



\begin{figure*}[t]
\vspace{-10pt}
\centering
\begin{subfigure}[b]{0.24\textwidth}
    \centering
    \includegraphics[width=\textwidth]{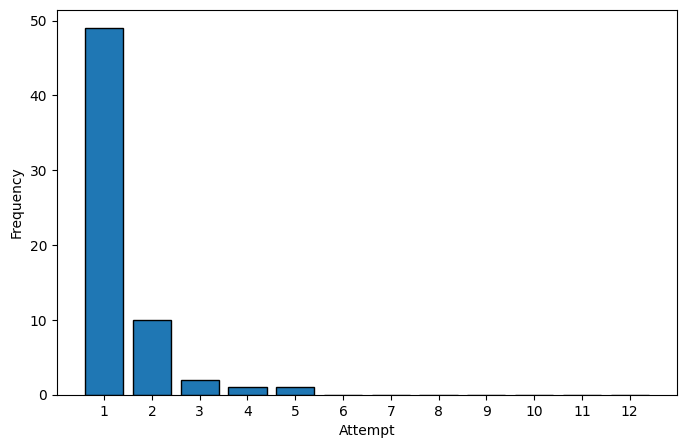}
    \vspace{-18pt}
    \caption{Ret. Val. Type/Range}
    \label{fig:a}
\end{subfigure}
\hfill
\begin{subfigure}[b]{0.24\textwidth}
    \centering
    \includegraphics[width=\textwidth]{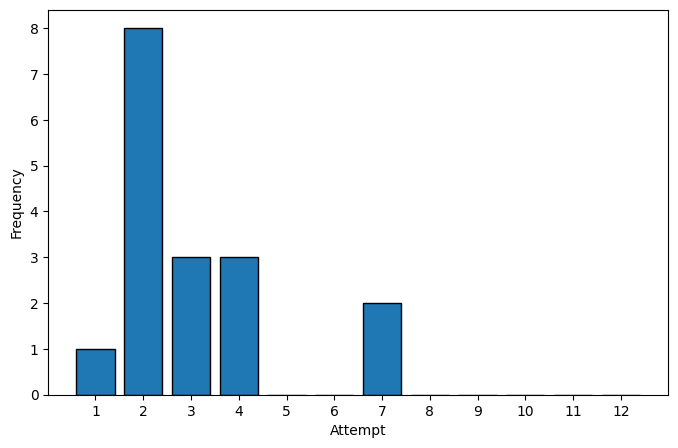}
    \vspace{-18pt}
    \caption{Check Base Cases}
    \label{fig:b}
\end{subfigure}
\hfill
\begin{subfigure}[b]{0.24\textwidth}
    \centering
    \includegraphics[width=\textwidth]{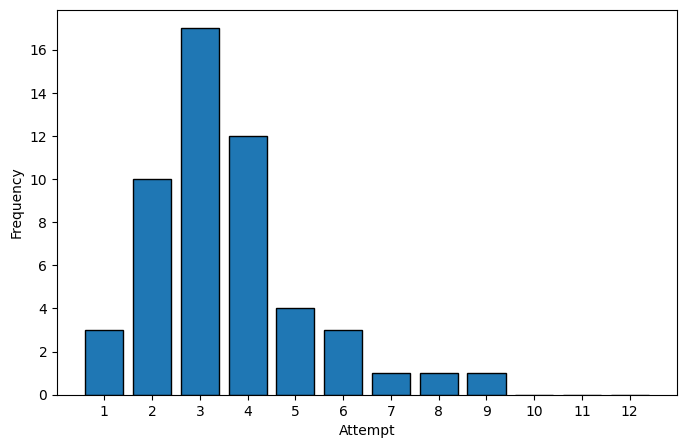}
    \vspace{-18pt}
    \caption{Check Edge Cases}
    \label{fig:c}
\end{subfigure}
\hfill
\begin{subfigure}[b]{0.24\textwidth}
    \centering
    \includegraphics[width=\textwidth]{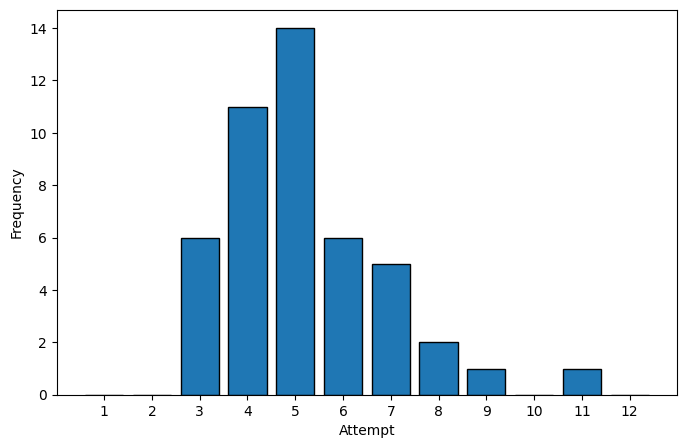}
    \vspace{-18pt}
    \caption{Postcond. Comb.}
    \label{fig:d}
\end{subfigure}
\vspace{12pt}
\begin{subfigure}[b]{0.27\textwidth}
    \centering
    \includegraphics[width=\textwidth]{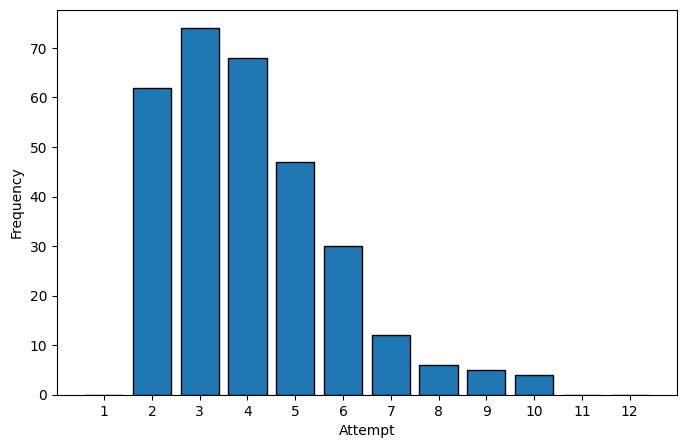}
    \vspace{-18pt}
    \caption{Refine Assertions}
    \label{fig:e}
\end{subfigure}
\hfill
\begin{subfigure}[b]{0.27\textwidth}
    \centering
    \includegraphics[width=\textwidth]{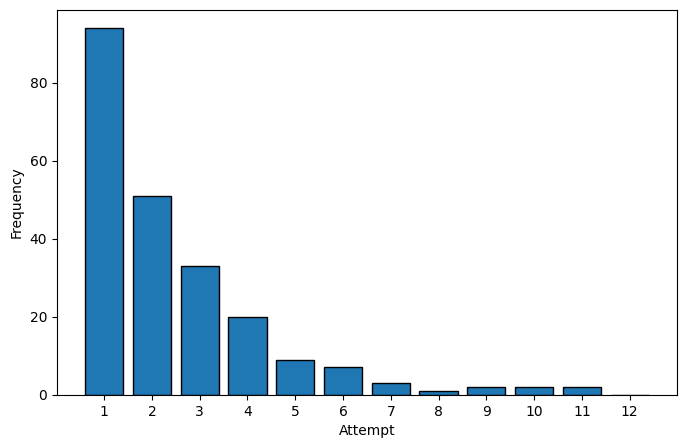}
    \vspace{-18pt}
    \caption{Verify Function Behavior}
    \label{fig:f}
\end{subfigure}
\hfill
\begin{subfigure}[b]{0.27\textwidth}
    \centering
    \includegraphics[width=\textwidth]{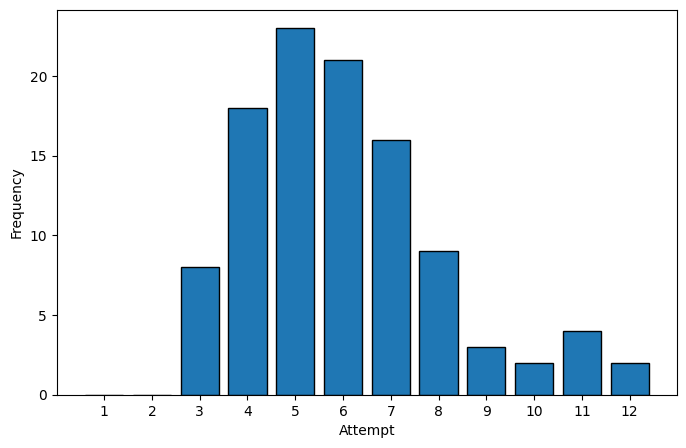}
    \vspace{-18pt}
    \caption{Submit A Solution}
    \label{fig:g}
\end{subfigure}
\vspace{-18pt}
\caption{Frequency distribution analysis from seven reasoning categories across attempts (RQ2)}
\label{fig:frequency_analysis}
\end{figure*}

\vspace{2pt}
{\bf Reasoning Patterns.} Fig.~\ref{fig:frequency_analysis} shows the~occurrences of {\em 7 categories across attempts} for all samples, excluding unsolvable~cases. 
\textit{Return Value Type and Range} and \textit{Verify Function Behavior} concentrate in the first attempt (49 and 94 instances). \textit{Check Base Cases} and \textit{Check Edge Cases} also appear early, peaking at the 2nd (8) and 3rd (17) attempts, indicating that the model initially explores return values and base/edge cases. \textit{Submit Final Solution} occurs from attempts 3–12 and peaks around attempts 5–6, consistent with Exploratory requiring about 5–6 attempts per case. \textit{Refine Assertions} is frequent between attempts 2–6, peaking at attempts 3–4 (74 and 68), suggesting the model refines before submitting. \textit{Postcondition Combination} emerges~at attempt 3 and peaks at attempt 5 (14), indicating it combines exploratory results before submitting. {\color{black}{See Appendix~\ref{sec:appendix_reasoning_example} for an example.
}}

\subsection{Stratifying Results on Hard Cases (RQ3)}

We analyze the hardest cases: the model exhausts the attempt budget without reaching the completeness threshold, under the strict setting $\tau{=}90,\mu{=}12$. This yields 23 hard cases out of 159: one with no correct postcondition and 22 with completeness $<90$ (average 56.2\%).
In the standard setting, we provide only binary feedback: whether a postcondition passes all tests (correctness) and whether~it catches all mutants (completeness). For these hard cases, we {\bf add richer feedback} by returning a randomly-selected uncaught mutant and asking the model for refinement. We then extend the budget by 4 attempts ($\mu{=}16$) to measure the~impact.

With enhanced feedback, correct postconditions increase to \textbf{22/23} and average completeness rises to \textbf{85.1\%}; the model reaches the threshold in 14/22 correct cases. Applying binary feedback~first ($\tau{=}90,\mu{=}12$) and using enhanced feedback only for the remaining hard cases with 4 extra attempts improves performance to \textbf{99.7\%} correctness and \textbf{92.0\%} completeness (vs. 99.4\% and 89.6\% with binary feedback only).
If we use enhanced feedback from the start with $\tau{=}90,\mu{=}12$, the model produces correct postconditions for 21/23 hard cases; among 22 correct cases, 9 reach the completeness target, with overall average completeness 90.6\%.

\subsection{\normalsize LLM Token Usage \& Cost Analysis (RQ4)}
\label{sec:tokencost}

Under the strictest configuration (completeness threshold $\tau$=90, maximum number of turns $\mu$=12), Random Sampling uses an average of 302 tokens per instance (min=19 and max=1,666 tokens), which is much lower than the usages of Multi-turn approaches.
Greedy uses an average of 5,752 tokens per instance (ranging from 266 to 25,909), while Exploratory  uses an average of 8,299 tokens per instance (ranging 1,313--29,956). 

The higher token usage per instance for our tool stems from two factors: (1) inclusion of Chain-of-Thought tokens (\code{nl2postcond} does not have CoT) and (2) inclusion of the history buffer (Equation~\ref{eq1}). However, the actual costs remain modest: 0.18 cent and 0.25 cent per instance for Greedy and Exploratory, respectively ($\tau$=90, $\mu$=12).

\begin{figure}[t]
    \centering    
    \includegraphics[width=0.45\textwidth]{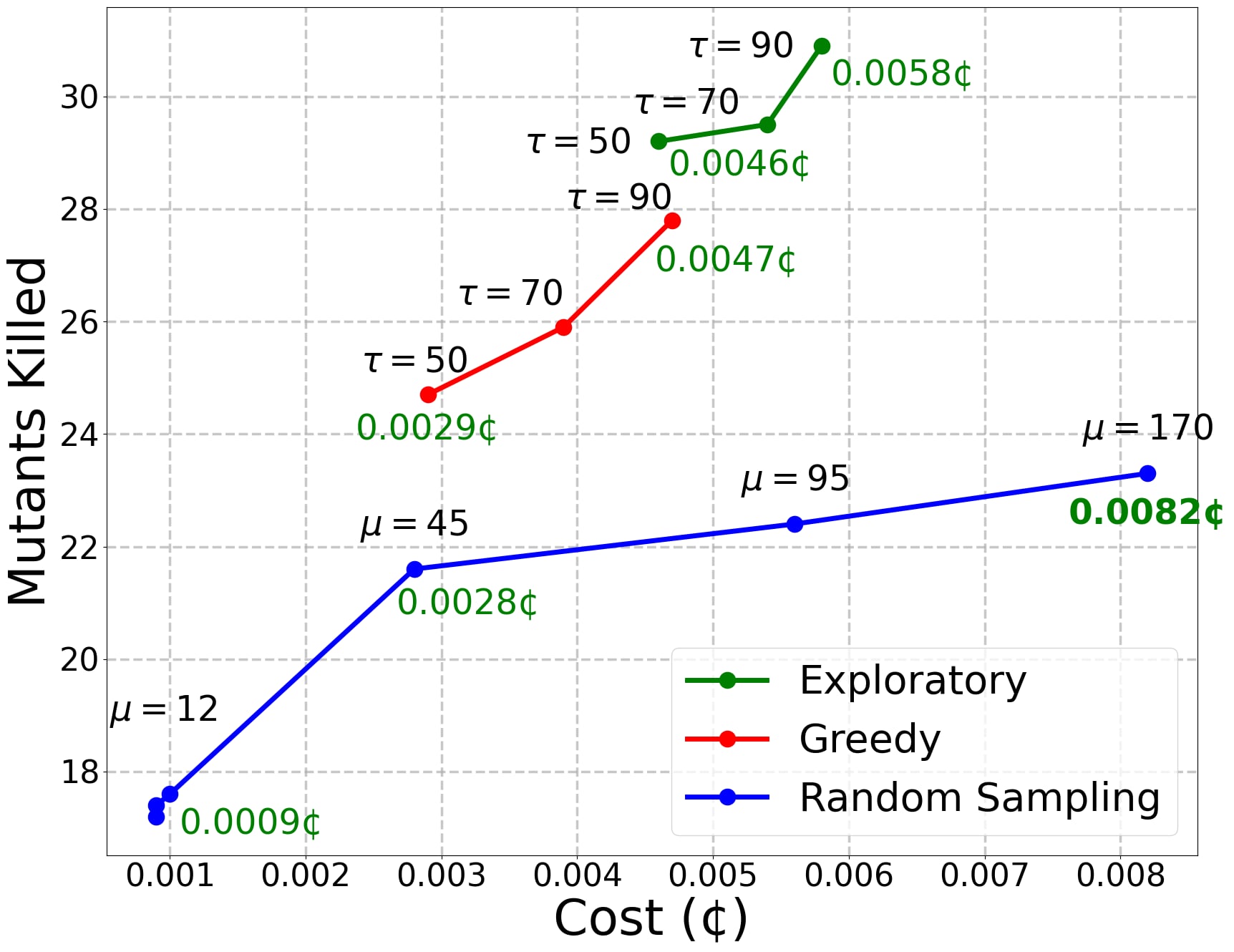} 
    \vspace{-8pt}
    \caption{Token Usages $\&$ Numbers of Caught Mutants}
    \label{fig:token_efficiency}
\end{figure}


This high token usage is a {\bf tradeoff for more bug detection}. Fig.~\ref{fig:token_efficiency} displays the trends for Random Sampling, Greedy, and Exploratory when the completeness threshold increases ($\mu$=50,70,90). When $\mu$ increases, each approach uses more tokens due to more attempts. However, across thresholds, both Greedy and Exploratory catch more mutants/bugs than Random Sampling. Greedy caught 24.7 mutants with 3,065.9 tokens per instance ($\approx$ 0.0029 cent/bug) and 27.8 mutants with 5,752 tokens per instance ($\approx$ 0.0047 cent/bug). Exploratory  achieved the highest bug detection performance, reaching 29.2 mutants with 6110.5 tokens per instance ($\approx$ 0.0046 cent/bug) and peaking at 30.9 with 8,299 tokens per instances ($\approx$ 0.0058 cent/bug). Random Sampling caught 17.6 mutants with 298 tokens. Using more tokens (3,182 tokens) with maximum allowed attempts $\mu$=170, its number of detected bugs reaches to 23.3 ({\bf 1.4 bugs less} compared to 24.7 bugs by Greedy with 3k tokens), with 0.0082 cent/bug ({\bf 2.83X cost more}). Random Sampling has higher cost as it generates more output tokens (which cost more), whereas our tool requires more input tokens but produces fewer outputs. In brief, {\bf with lower attempts, Exploratory incurs lower costs with more detected~bugs}.





\subsection{{\color{custom-blue}{Usefulness in Bug Detection  (RQ5)}}}
\label{sec:appendix_bug}


{\bf Experimental Methodology}. 
We adapted an evaluation approach used in \code{nl2postcond}~\cite{nl2postcond} and TOGA~\cite{dinella2022toga}. The evaluation idea is that it aims to evaluate if a model can generate postconditions that distinguish between correct and buggy code regarding regression and triggering test cases in regression testing. 
We used FixEval~\cite{fixeval}, a benchmark comprising buggy and correct submission code to competitive programming problems. From the full test set of 243k Python buggy-correct pairs, we filtered out runtime errors and kept 157,355 pairs. We then 
randomly selected 384 bugs, ensuring 95\% statistical significance. We evaluated the effectiveness of postconditions that a user could have used to catch a bug as the postconditions fail on the buggy version, and succeed on the correct one. 
The rationale is as follows. Using the associated test suite for a problem in the dataset to ensure correctness, as we run a model on the correct version, the resulting postcondition $\phi$ acts as an oracle of intended behavior of the code. When we apply that oracle to the executions of the buggy solution, any deviation from the intended relation between inputs and outputs will violate the postcondition $\phi$ (i.e., there exists an input $i$ such that postcondition $\phi$(i,buggy\_function(i))=\code{false}). This is the same completeness definition: postconditions ``reject'' a faulty variant (mutant); here, the buggy version plays the role of a single mutant. Thus, the generated postcondition can be used in detecting future bugs in regression~testing.
We used the same criteria in \code{nl2postcond} for a {\em bug-discriminating} postcondition. We consider a generated postcondition to be bug-discriminating if it satisfies the following:
(1) The postcondition passes all the trigger and regression tests, on the correct version of a function. (2) The postcondition fails a trigger test or a regression test on the buggy version~\cite{nl2postcond}. 


\begin{table}[t]
\centering
\small
\tabcolsep 1.5pt
\caption{Bug Detection Performance.}
\vspace{-6pt}
\label{tab:bug}
\begin{tabular}{@{}l|cc|cc|cc@{}}
\toprule
\multicolumn{1}{c|}{\multirow{2}{*}{Approach}} & \multicolumn{2}{c|}{$\mu = 3$} & \multicolumn{2}{c|}{$\mu = 5$} & \multicolumn{2}{c}{$\mu = 10$} \\ \cmidrule(l){2-7} 
\multicolumn{1}{c|}{} 
& \multicolumn{1}{c}{\% Corr.} 
& \multicolumn{1}{c|}{\begin{tabular}[c]{@{}c@{}}\% dist.\\ bugs\end{tabular}} 
& \multicolumn{1}{c}{\% Corr.} 
& \multicolumn{1}{c|}{\begin{tabular}[c]{@{}c@{}}\% dist.\\ bugs\end{tabular}} 
& \multicolumn{1}{c}{\% Corr.} 
& \multicolumn{1}{c}{\begin{tabular}[c]{@{}c@{}}\% dist.\\ bugs\end{tabular}} \\ 
\midrule
R.Sampl. w/ & \multirow{2}{*}{66.41} & \multirow{2}{*}{16.15} & \multirow{2}{*}{76.82} & \multirow{2}{*}{22.40} & \multirow{2}{*}{86.20} & \multirow{2}{*}{27.86} \\
nl2postcond & & & & & & \\
Greedy & 59.38 & 24.74 & 67.71 & 30.73 & 87.24 & 38.02 \\
Exploratory & {\bf 79.69} & {\bf 34.64} & {\bf 90.10} & {\bf 36.98} & {\bf 95.57} & {\bf 39.58} \\
\bottomrule
\end{tabular}
\end{table}

\noindent {\bf Experimental Results}. As seen in Table~\ref{tab:bug}, both Greedy and Exploratory approaches are able to distinguish more buggy versions than the baseline \code{nl2postcond} being run $\mu$ times independently (for a fair comparison). When $\mu$=3 submission attempts, the baseline distinguished 16.15\% of the buggy versions from the correct ones, with passing 66.41\% of test cases. In contrast, Greedy distinguishes relatively 53.19\% more buggy versions with passing 59.38\% of test cases. Importantly, Exploratory catches {\bf 2X} more buggy versions than the baseline with {\bf 34.64\%} of passing test cases.

As we increased the number of submission attempts to $\mu$=5, all approaches performed better in terms of both correctness and number of distinguishable bugs. As $\mu$=10, Exploratory performs the best as it distinguished {\bf 39.58\%} of the buggy versions, i.e., better than both the baseline and Greedy. 

An example of the bugs detected by the postconditions from {\tool} is shown in Section~\ref{appendix:bug-example}.

\section*{Limitations \& Threats to Validity}
\label{sec:threats}

Our evaluation is conducted on EvalPlus~\cite{nl2postcond}, which may not reflect the full diversity of real-world codebases. While EvalPlus is a large public benchmark with postcondition annotations and enables a fair comparison against \code{nl2postcond} on its original setting, it primarily consists of Python tasks of moderate complexity. Thus, generalization to larger systems, other programming languages, or other forms of specifications is not guaranteed. In addition, we evaluate mainly GPT-family models; performance may differ for other model families or future versions.

Our results may also be influenced by the pre-training knowledge of LLMs. Although we cannot fully rule out exposure to similar code during pretraining, our evaluation centers on generating new postconditions and validating them via execution, which is unlikely to be explained by memorization.

We operationalize postcondition quality using two proxies: correctness (passing the test suite) and completeness (rejecting mutants). Both  inherit limitations from their backends: correctness is bounded by test coverage, and completeness depends on the quality and representativeness of the mutant set. If the tests or mutants fail to capture realistic behaviors or defects, our metrics may overestimate true specification quality or bug-detection usefulness.

Finally, our conclusions depend on experimental choices such as the attempt budget $\mu$ and completeness threshold $\tau$. While we evaluate across multiple configurations to reduce sensitivity to any single setting, other datasets, feedback designs, or hyperparameter choices could change the observed trade-offs between effectiveness and efficiency.

\bibliography{references, refs-specmining}

@article{chen2021evaluating,
  title={Evaluating large language models trained on code},
  author={Chen, Mark and Tworek, Jerry and Jun, Heewoo and Yuan, Qiming and Pinto, Henrique Ponde De Oliveira and Kaplan, Jared and Edwards, Harri and Burda, Yuri and Joseph, Nicholas and Brockman, Greg and others},
  journal={arXiv preprint arXiv:2107.03374},
  year={2021}
}

@article{nl2postcond,
author = {Endres, Madeline and Fakhoury, Sarah and Chakraborty, Saikat and Lahiri, Shuvendu K.},
title = {Can Large Language Models Transform Natural Language Intent into Formal Method Postconditions?},
year = {2024},
issue_date = {July 2024},
publisher = {Association for Computing Machinery},
address = {New York, NY, USA},
volume = {1},
number = {FSE},
url = {https://doi.org/10.1145/3660791},
doi = {10.1145/3660791},
journal = {Proc. ACM Softw. Eng.},
month = jul,
articleno = {84},
numpages = {24}
}

@inproceedings{
gehring2025rlef,
title={{RLEF}: Grounding Code {LLM}s in Execution Feedback with Reinforcement Learning},
author={Jonas Gehring and Kunhao Zheng and Jade Copet and Vegard Mella and Taco Cohen and Gabriel Synnaeve},
booktitle={Forty-second International Conference on Machine Learning},
year={2025}
}

@inproceedings{evalplus,
  title = {Is Your Code Generated by Chat{GPT} Really Correct? Rigorous Evaluation of Large Language Models for Code Generation},
  author = {Liu, Jiawei and Xia, Chunqiu Steven and Wang, Yuyao and Zhang, Lingming},
  booktitle = {Thirty-seventh Conference on Neural Information Processing Systems},
  year = {2023},
  url = {https://openreview.net/forum?id=1qvx610Cu7},
}

@inproceedings{10.1145/1273463.1273487,
author = {Shoham, Sharon and Yahav, Eran and Fink, Stephen and Pistoia, Marco},
title = {Static specification mining using automata-based abstractions},
year = {2007},
isbn = {9781595937346},
publisher = {Association for Computing Machinery},
address = {New York, NY, USA},
url = {https://doi.org/10.1145/1273463.1273487},
doi = {10.1145/1273463.1273487},
booktitle = {Proceedings of the 2007 International Symposium on Software Testing and Analysis},
pages = {174–184},
numpages = {11},
location = {London, United Kingdom},
series = {ISSTA '07}
}

@inproceedings{10.1145/2384616.2384633,
author = {Cousot, Patrick M. and Cousot, Radhia and Logozzo, Francesco and Barnett, Michael},
title = {An abstract interpretation framework for refactoring with application to extract methods with contracts},
year = {2012},
isbn = {9781450315616},
publisher = {Association for Computing Machinery},
address = {New York, NY, USA},
url = {https://doi.org/10.1145/2384616.2384633},
doi = {10.1145/2384616.2384633},
booktitle = {Proceedings of the ACM International Conference on Object Oriented Programming Systems Languages and Applications},
pages = {213–232},
numpages = {20},
location = {Tucson, Arizona, USA},
series = {OOPSLA '12}
}

@inproceedings{10.5555/2337223.2337319,
author = {Pandita, Rahul and Xiao, Xusheng and Zhong, Hao and Xie, Tao and Oney, Stephen and Paradkar, Amit},
title = {Inferring method specifications from natural language API descriptions},
year = {2012},
isbn = {9781467310673},
publisher = {IEEE Press},
booktitle = {Proceedings of the 34th International Conference on Software Engineering},
pages = {815–825},
numpages = {11},
location = {Zurich, Switzerland},
series = {ICSE '12}
}

@inproceedings{10.1145/1294261.1294276,
author = {Tan, Lin and Yuan, Ding and Krishna, Gopal and Zhou, Yuanyuan},
title = {/*icomment: bugs or bad comments?*/},
year = {2007},
isbn = {9781595935915},
publisher = {Association for Computing Machinery},
address = {New York, NY, USA},
url = {https://doi.org/10.1145/1294261.1294276},
doi = {10.1145/1294261.1294276},
booktitle = {Proceedings of Twenty-First ACM SIGOPS Symposium on Operating Systems Principles},
pages = {145–158},
numpages = {14},
location = {Stevenson, Washington, USA},
series = {SOSP '07}
}

@inproceedings{10.1145/1985793.1985796,
author = {Tan, Lin and Zhou, Yuanyuan and Padioleau, Yoann},
title = {aComment: mining annotations from comments and code to detect interrupt related concurrency bugs},
year = {2011},
isbn = {9781450304450},
publisher = {Association for Computing Machinery},
address = {New York, NY, USA},
url = {https://doi.org/10.1145/1985793.1985796},
doi = {10.1145/1985793.1985796},
booktitle = {Proceedings of the 33rd International Conference on Software Engineering},
pages = {11–20},
numpages = {10},
location = {Waikiki, Honolulu, HI, USA},
series = {ICSE '11}
}

@inproceedings{10.1109/ICST.2012.106,
author = {Tan, Shin Hwei and Marinov, Darko and Tan, Lin and Leavens, Gary T.},
title = {@tComment: Testing Javadoc Comments to Detect Comment-Code Inconsistencies},
year = {2012},
isbn = {9780769546704},
publisher = {IEEE Computer Society},
address = {USA},
url = {https://doi.org/10.1109/ICST.2012.106},
doi = {10.1109/ICST.2012.106},
booktitle = {Proceedings of the 2012 IEEE Fifth International Conference on Software Testing, Verification and Validation},
pages = {260–269},
numpages = {10},
series = {ICST '12}
}

@inproceedings{10.1145/3213846.3213872,
author = {Blasi, Arianna and Goffi, Alberto and Kuznetsov, Konstantin and Gorla, Alessandra and Ernst, Michael D. and Pezz\`{e}, Mauro and Castellanos, Sergio Delgado},
title = {Translating code comments to procedure specifications},
year = {2018},
isbn = {9781450356992},
publisher = {Association for Computing Machinery},
address = {New York, NY, USA},
url = {https://doi.org/10.1145/3213846.3213872},
doi = {10.1145/3213846.3213872},
booktitle = {Proceedings of the 27th ACM SIGSOFT International Symposium on Software Testing and Analysis},
pages = {242–253},
numpages = {12},
location = {Amsterdam, Netherlands},
series = {ISSTA 2018}
}

@inproceedings{10.1109/ASE.2009.94,
author = {Zhong, Hao and Zhang, Lu and Xie, Tao and Mei, Hong},
title = {Inferring Resource Specifications from Natural Language API Documentation},
year = {2009},
isbn = {9780769538914},
publisher = {IEEE Computer Society},
address = {USA},
url = {https://doi.org/10.1109/ASE.2009.94},
doi = {10.1109/ASE.2009.94},
booktitle = {Proceedings of the 24th IEEE/ACM International Conference on Automated Software Engineering},
pages = {307–318},
numpages = {12},
series = {ASE '09}
}

@INPROCEEDINGS{7985647,
  author={Zhou, Yu and Gu, Ruihang and Chen, Taolue and Huang, Zhiqiu and Panichella, Sebastiano and Gall, Harald},
  booktitle={2017 IEEE/ACM 39th International Conference on Software Engineering (ICSE)}, 
  title={Analyzing APIs Documentation and Code to Detect Directive Defects}, 
  year={2017},
  volume={},
  number={},
  pages={27-37},
  doi={10.1109/ICSE.2017.11}}

@inproceedings{dinella2022toga,
author = {Dinella, Elizabeth and Ryan, Gabriel and Mytkowicz, Todd and Lahiri, Shuvendu K.},
title = {TOGA: a neural method for test oracle generation},
year = {2022},
isbn = {9781450392211},
publisher = {Association for Computing Machinery},
address = {New York, NY, USA},
url = {https://doi.org/10.1145/3510003.3510141},
doi = {10.1145/3510003.3510141},
booktitle = {Proceedings of the 44th International Conference on Software Engineering},
pages = {2130–2141},
numpages = {12},
location = {Pittsburgh, Pennsylvania},
series = {ICSE '22}
}

@ARTICLE{mastropaolo2023usingtransfer,
  author={Mastropaolo, Antonio and Cooper, Nathan and Palacio, David Nader and Scalabrino, Simone and Poshyvanyk, Denys and Oliveto, Rocco and Bavota, Gabriele},
  journal={IEEE Transactions on Software Engineering}, 
  title={Using Transfer Learning for Code-Related Tasks}, 
  year={2023},
  volume={49},
  number={4},
  pages={1580-1598},
  doi={10.1109/TSE.2022.3183297}}

@inproceedings{10.1145/3524481.3527220,
author = {Tufano, Michele and Drain, Dawn and Svyatkovskiy, Alexey and Sundaresan, Neel},
title = {Generating accurate assert statements for unit test cases using pretrained transformers},
year = {2022},
isbn = {9781450392860},
publisher = {Association for Computing Machinery},
address = {New York, NY, USA},
url = {https://doi.org/10.1145/3524481.3527220},
doi = {10.1145/3524481.3527220},
booktitle = {Proceedings of the 3rd ACM/IEEE International Conference on Automation of Software Test},
pages = {54–64},
numpages = {11},
location = {Pittsburgh, Pennsylvania},
series = {AST '22}
}

@inproceedings{lemieux2023codamosa,
author = {Lemieux, Caroline and Inala, Jeevana Priya and Lahiri, Shuvendu K. and Sen, Siddhartha},
title = {CodaMosa: Escaping Coverage Plateaus in Test Generation with Pre-Trained Large Language Models},
year = {2023},
isbn = {9781665457019},
publisher = {IEEE Press},
url = {https://doi.org/10.1109/ICSE48619.2023.00085},
doi = {10.1109/ICSE48619.2023.00085},
booktitle = {Proceedings of the 45th International Conference on Software Engineering},
pages = {919–931},
numpages = {13},
location = {Melbourne, Victoria, Australia},
series = {ICSE '23}
}

@article{coverup2025,
author = {Altmayer Pizzorno, Juan and Berger, Emery D.},
title = {CoverUp: Effective High Coverage Test Generation for Python},
year = {2025},
issue_date = {July 2025},
publisher = {Association for Computing Machinery},
address = {New York, NY, USA},
volume = {2},
number = {FSE},
url = {https://doi.org/10.1145/3729398},
doi = {10.1145/3729398},
journal = {Proc. ACM Softw. Eng.},
month = jun,
articleno = {FSE128},
numpages = {23}
}

@misc{lahiri2023interactivecodegenerationtestdriven,
      title={Interactive Code Generation via Test-Driven User-Intent Formalization}, 
      author={Shuvendu K. Lahiri and Sarah Fakhoury and Aaditya Naik and Georgios Sakkas and Saikat Chakraborty and Madanlal Musuvathi and Piali Choudhury and Curtis von Veh and Jeevana Priya Inala and Chenglong Wang and Jianfeng Gao},
      year={2023},
      eprint={2208.05950},
      archivePrefix={arXiv},
      primaryClass={cs.SE},
      url={https://arxiv.org/abs/2208.05950}, 
}

@misc{tufano2021unittestcasegeneration,
      title={Unit Test Case Generation with Transformers and Focal Context}, 
      author={Michele Tufano and Dawn Drain and Alexey Svyatkovskiy and Shao Kun Deng and Neel Sundaresan},
      year={2021},
      eprint={2009.05617},
      archivePrefix={arXiv},
      primaryClass={cs.SE},
      url={https://arxiv.org/abs/2009.05617}, 
}

@inproceedings{10.1109/ICSE43902.2021.00112,
author = {Molina, Facundo and Ponzio, Pablo and Aguirre, Nazareno and Frias, Marcelo},
title = {EvoSpex: An Evolutionary Algorithm for Learning Postconditions},
year = {2021},
isbn = {9781450390859},
publisher = {IEEE Press},
url = {https://doi.org/10.1109/ICSE43902.2021.00112},
doi = {10.1109/ICSE43902.2021.00112},
booktitle = {Proceedings of the 43rd International Conference on Software Engineering},
pages = {1223–1235},
numpages = {13},
location = {Madrid, Spain},
series = {ICSE '21}
}

@misc{vikram2024largelanguagemodelswrite,
      title={Can Large Language Models Write Good Property-Based Tests?}, 
      author={Vasudev Vikram and Caroline Lemieux and Joshua Sunshine and Rohan Padhye},
      year={2024},
      eprint={2307.04346},
      archivePrefix={arXiv},
      primaryClass={cs.SE},
      url={https://arxiv.org/abs/2307.04346}, 
}

@article{DBLP:journals/corr/abs-2210-00848,
  author       = {Darren Key and
                  Wen{-}Ding Li and
                  Kevin Ellis},
  title        = {I Speak, You Verify: Toward Trustworthy Neural Program Synthesis},
  journal      = {CoRR},
  volume       = {abs/2210.00848},
  year         = {2022},
  url          = {https://doi.org/10.48550/arXiv.2210.00848},
  doi          = {10.48550/ARXIV.2210.00848},
  eprinttype    = {arXiv},
  eprint       = {2210.00848},
  bibsource    = {dblp computer science bibliography, https://dblp.org}
}

@inproceedings{10.1145/2837614.2837664,
author = {Garg, Pranav and Neider, Daniel and Madhusudan, P. and Roth, Dan},
title = {Learning invariants using decision trees and implication counterexamples},
year = {2016},
isbn = {9781450335492},
publisher = {Association for Computing Machinery},
address = {New York, NY, USA},
url = {https://doi.org/10.1145/2837614.2837664},
doi = {10.1145/2837614.2837664},
booktitle = {Proceedings of the 43rd Annual ACM SIGPLAN-SIGACT Symposium on Principles of Programming Languages},
pages = {499–512},
numpages = {14},
location = {St. Petersburg, FL, USA},
series = {POPL '16}
}

@inproceedings{
Laich2020Guiding,
title={Guiding Program Synthesis by Learning to Generate Examples},
author={Larissa Laich and Pavol Bielik and Martin Vechev},
booktitle={International Conference on Learning Representations},
year={2020},
url={https://openreview.net/forum?id=BJl07ySKvS}
}

@inproceedings{10.1145/3385412.3385986,
author = {Yao, Jianan and Ryan, Gabriel and Wong, Justin and Jana, Suman and Gu, Ronghui},
title = {Learning nonlinear loop invariants with gated continuous logic networks},
year = {2020},
isbn = {9781450376136},
publisher = {Association for Computing Machinery},
address = {New York, NY, USA},
url = {https://doi.org/10.1145/3385412.3385986},
doi = {10.1145/3385412.3385986},
booktitle = {Proceedings of the 41st ACM SIGPLAN Conference on Programming Language Design and Implementation},
pages = {106–120},
numpages = {15},
location = {London, UK},
series = {PLDI 2020}
}

@InProceedings{pmlr-v202-pei23a,
  title = 	 {Can Large Language Models Reason about Program Invariants?},
  author =       {Pei, Kexin and Bieber, David and Shi, Kensen and Sutton, Charles and Yin, Pengcheng},
  booktitle = 	 {Proceedings of the 40th International Conference on Machine Learning},
  pages = 	 {27496--27520},
  year = 	 {2023},
  editor = 	 {Krause, Andreas and Brunskill, Emma and Cho, Kyunghyun and Engelhardt, Barbara and Sabato, Sivan and Scarlett, Jonathan},
  volume = 	 {202},
  series = 	 {Proceedings of Machine Learning Research},
  month = 	 {23--29 Jul},
  publisher =    {PMLR},
  url = 	 {https://proceedings.mlr.press/v202/pei23a.html}
}

@INPROCEEDINGS {fixeval,
author = { Anjum Haque, Md Mahim and Ahmad, Wasi Uddin and Lourentzou, Ismini and Brown, Chris },
booktitle = { 2023 IEEE/ACM International Workshop on Automated Program Repair (APR) },
title = {{ FixEval: Execution-based Evaluation of Program Fixes for Programming Problems }},
year = {2023},
volume = {},
ISSN = {},
pages = {11-18},
doi = {10.1109/APR59189.2023.00009},
url = {https://doi.ieeecomputersociety.org/10.1109/APR59189.2023.00009},
publisher = {IEEE Computer Society},
address = {Los Alamitos, CA, USA},
month =May}

@misc{Specmind,
  author = {},
  title = {SpecMind Code},
  year = {2026},
  howpublished = {\url{https://github.com/thaiminhpv/SpecMind}},
}

@inproceedings{zeller07,
 author = {Wasylkowski, Andrzej and Zeller, Andreas and Lindig, Christian},
 title = {Detecting Object Usage Anomalies},
 booktitle = {Proceedings of the Symposium on Foundations of Software Engineering},
 series = {ESEC-FSE '07},
 year = {2007},
 isbn = {978-1-59593-811-4},
 location = {Dubrovnik, Croatia},
 pages = {35--44},
 numpages = {10},
 url = {http://doi.acm.org/10.1145/1287624.1287632},
 doi = {10.1145/1287624.1287632},
 acmid = {1287632},
 publisher = {ACM},
}

@inproceedings{dynamine05,
author = {Livshits, Benjamin and Zimmermann, Thomas},
title = {DynaMine: finding common error patterns by mining software revision histories},
year = {2005},
isbn = {1595930140},
publisher = {Association for Computing Machinery},
address = {New York, NY, USA},
url = {https://doi.org/10.1145/1081706.1081754},
doi = {10.1145/1081706.1081754},
booktitle = {Proceedings of the 10th European Software Engineering Conference Held Jointly with 13th ACM SIGSOFT International Symposium on Foundations of Software Engineering},
pages = {296–305},
numpages = {10},
location = {Lisbon, Portugal},
series = {ESEC/FSE-13}
}

@inproceedings{engler-sosp01,
 author = {Engler, Dawson and Chen, David Yu and Hallem, Seth and Chou, Andy and Chelf, Benjamin},
 title = {Bugs As Deviant Behavior: A General Approach to Inferring Errors in Systems Code},
 booktitle = {Proceedings of the Eighteenth ACM Symposium on Operating Systems Principles},
 series = {SOSP'01},
 year = {2001},
 isbn = {1-58113-389-8},
 location = {Banff, Alberta, Canada},
 pages = {57--72},
 numpages = {16},
 url = {http://doi.acm.org/10.1145/502034.502041},
 doi = {10.1145/502034.502041},
 acmid = {502041},
 publisher = {ACM},
}

@article{williams-tse05,
 author = {Williams,, Chadd C. and Hollingsworth,, Jeffrey K.},
 title = {Automatic Mining of Source Code Repositories to Improve Bug Finding Techniques},
 journal = {IEEE Trans. Softw. Eng.},
 volume = {31},
 number = {6},
 year = {2005},
 pages = {466--480},
 publisher = {IEEE Press},
}

@inproceedings{prminer-fse05,
 author = {Li, Zhenmin and Zhou, Yuanyuan},
 title = {PR-Miner: Automatically Extracting Implicit Programming Rules and Detecting Violations in Large Software Code},
 booktitle = {Proceedings of the 13th Symposium on Foundations of Software Engineering},
 series = {ESEC/FSE-13},
 year = {2005},
 isbn = {1-59593-014-0},
 location = {Lisbon, Portugal},
 pages = {306--315},
 numpages = {10},
 url = {http://doi.acm.org/10.1145/1081706.1081755},
 doi = {10.1145/1081706.1081755},
 acmid = {1081755},
 publisher = {ACM},
}

@inproceedings{ammons-popl02,
 author = {Ammons, Glenn and Bod\'{\i}k, Rastislav and Larus, James R.},
 title = {Mining Specifications},
 booktitle = {Proceedings of the 29th ACM SIGPLAN SIGACT Symposium on Principles of Programming Languages},
 series = {POPL '02},
 year = {2002},
 isbn = {1-58113-450-9},
 location = {Portland, Oregon},
 pages = {4--16},
 numpages = {13},
 url = {http://doi.acm.org/10.1145/503272.503275},
 doi = {10.1145/503272.503275},
 acmid = {503275},
 publisher = {ACM},
}

@INPROCEEDINGS{taoxie-ase09,
 author = {Thummalapenta, Suresh and Xie, Tao},
 title = {Alattin: Mining Alternative Patterns for Detecting Neglected Conditions},
 booktitle = {Proceedings of the 2009 IEEE/ACM International Conference on Automated Software Engineering},
 series = {ASE '09},
 year = {2009},
 isbn = {978-0-7695-3891-4},
 pages = {283--294},
 numpages = {12},
 url = {http://dx.doi.org/10.1109/ASE.2009.72},
 doi = {10.1109/ASE.2009.72},
 acmid = {1747526},
 publisher = {IEEE Computer Society},
}

@INPROCEEDINGS{zeller-ase09,
 author = {Wasylkowski, Andrzej and Zeller, Andreas},
 title = {Mining Temporal Specifications from Object Usage},
 booktitle = {Proceedings of the 2009 IEEE/ACM International Conference on Automated Software Engineering},
 series = {ASE '09},
 year = {2009},
 isbn = {978-0-7695-3891-4},
 pages = {295--306},
 numpages = {12},
 url = {http://dx.doi.org/10.1109/ASE.2009.30},
 doi = {10.1109/ASE.2009.30},
 acmid = {1747527},
 publisher = {IEEE Computer Society},
}

@INPROCEEDINGS{mike-ase09,
 author = {Pradel, Michael and Gross, Thomas R.},
 title = {Automatic Generation of Object Usage Specifications from Large Method Traces},
 booktitle = {Proceedings of the 2009 IEEE/ACM International Conference on Automated Software Engineering},
 series = {ASE '09},
 year = {2009},
 isbn = {978-0-7695-3891-4},
 pages = {371--382},
 numpages = {12},
 url = {http://dx.doi.org/10.1109/ASE.2009.60},
 doi = {10.1109/ASE.2009.60},
 acmid = {1747533},
 publisher = {IEEE Computer Society},
}

@inproceedings{ramanathan-pldi07,
 author = {Ramanathan, Murali Krishna and Grama, Ananth and Jagannathan, Suresh},
 title = {Static Specification Inference Using Predicate Mining},
 booktitle = {Proceedings of the 2007 ACM SIGPLAN Conference on Programming Language Design and Implementation},
 series = {PLDI '07},
 year = {2007},
 isbn = {978-1-59593-633-2},
 location = {San Diego, California, USA},
 pages = {123--134},
 numpages = {12},
 url = {http://doi.acm.org/10.1145/1250734.1250749},
 doi = {10.1145/1250734.1250749},
 acmid = {1250749},
 publisher = {ACM},
}

@inproceedings{fse09,
 author = {Nguyen, Tung Thanh and Nguyen, Hoan Anh and Pham, Nam H. and Al-Kofahi, Jafar M. and Nguyen, Tien N.},
 title = {Graph-based Mining of Multiple Object Usage Patterns},
 booktitle = {Proceedings of the Symposium on Foundations of Software Engineering},
 series = {ESEC/FSE '09},
 year = {2009},
 isbn = {978-1-60558-001-2},
 location = {Amsterdam, The Netherlands},
 pages = {383--392},
 numpages = {10},
 url = {http://doi.acm.org/10.1145/1595696.1595767},
 doi = {10.1145/1595696.1595767},
 acmid = {1595767},
 publisher = {ACM},
}

@inproceedings{mapo09,
 author = {Zhong, Hao and Xie, Tao and Zhang, Lu and Pei, Jian and Mei, Hong},
 title = {MAPO: Mining and Recommending API Usage Patterns},
 booktitle = {Proceedings of the 23rd European Conference on ECOOP 2009 --- Object-Oriented Programming},
 year = {2009},
 pages = {318--343},
 publisher = {Springer-Verlag},
}

@inproceedings{kremenek06,
author = {Kremenek, Ted and Twohey, Paul and Back, Godmar and Ng, Andrew and Engler, Dawson},
title = {From uncertainty to belief: inferring the specification within},
year = {2006},
isbn = {1931971471},
publisher = {USENIX Association},
address = {USA},
booktitle = {Proceedings of the 7th Symposium on Operating Systems Design and Implementation},
pages = {161–176},
numpages = {16},
location = {Seattle, Washington},
series = {OSDI '06}
}

@inproceedings{daikon99,
 author = {Ernst, Michael D. and Cockrell, Jake and Griswold, William G. and Notkin, David},
 title = {Dynamically Discovering Likely Program Invariants to Support Program Evolution},
 booktitle = {Proceedings of the 21st International Conference on Software Engineering},
 series = {ICSE'99},
 year = {1999},
 isbn = {1-58113-074-0},
 location = {Los Angeles, California, USA},
 pages = {213--224},
 numpages = {12},
 url = {http://doi.acm.org/10.1145/302405.302467},
 doi = {10.1145/302405.302467},
 acmid = {302467},
 publisher = {ACM},
}

@inproceedings{yiwei-icse11,
 author = {Wei, Yi and Furia, Carlo A. and Kazmin, Nikolay and Meyer, Bertrand},
 title = {Inferring Better Contracts},
 booktitle = {Proceedings of the 33rd International Conference on Software Engineering},
 series = {ICSE '11},
 year = {2011},
 isbn = {978-1-4503-0445-0},
 location = {Waikiki, Honolulu, HI, USA},
 pages = {191--200},
 numpages = {10},
 url = {http://doi.acm.org/10.1145/1985793.1985820},
 doi = {10.1145/1985793.1985820},
 acmid = {1985820},
 publisher = {ACM},
}

@inproceedings{BeschastnikhBSSE2011,
 author = {Beschastnikh, Ivan and Brun, Yuriy and Schneider, Sigurd and Sloan, Michael and Ernst, Michael D.},
 title = {Leveraging Existing Instrumentation to Automatically Infer Invariant-constrained Models},
 booktitle = {Proceedings of the 19th Symposium on Foundations of Software Engineering},
 series = {ESEC/FSE '11},
 year = {2011},
 isbn = {978-1-4503-0443-6},
 location = {Szeged, Hungary},
 pages = {267--277},
 numpages = {11},
 url = {http://doi.acm.org/10.1145/2025113.2025151},
 doi = {10.1145/2025113.2025151},
 acmid = {2025151},
 publisher = {ACM},
}

@inproceedings{le2026testweaver,
  author        = {Le, Cuong Chi and Van, Cuong Duc and Vu, Tung Duy and Pham, Minh V. T. and Phan, Hoang Nhat and Phan, Huy N. and Nguyen, Tien N.},
  title         = {{TestWeaver}: Execution-aware, Feedback-driven Regression Testing Generation with Large Language Models},
  booktitle     = {Proceedings of the 48th IEEE/ACM International Conference on Software Engineering (ICSE '26)},
  year          = {2026},
  address       = {Rio de Janeiro, Brazil},
  month         = apr,
  note          = {To appear},
  url           = {https://arxiv.org/abs/2508.01255},
  eprint        = {2508.01255},
  archivePrefix = {arXiv},
  primaryClass  = {cs.SE}
}

\appendix
\section{Appendix}
\label{sec:appendix}

\subsection{Results of {\tool} with GLM-4.7}
\label{sec:appendix_glm}
\begin{table*}[t]
\vspace{-6pt}
\scriptsize
\centering
\caption{
{\color{custom-blue}{Postcondition Generation Effectiveness with GLM-4.7 (RQ1). $\tau$: completeness threshold, $\mu$: max turns; R. Sampl.: run nl2postcond $\mu$ independent times, Subs: avg submissions, Corr: correctness, Comp.: Completeness.}}
}
\vspace{-6pt}
\tabcolsep 1.7pt
\label{tab:rq1-glm}
\begin{minipage}{0.49\textwidth}

\begin{tabular}{l|c|c|c|c|c|}
\toprule
\textbf{Method} & \textbf{Config.} & \textbf{Attempts} & \textbf{Avg Subs} & \textbf{Corr.} & \textbf{Comp.} \\
 & & \textbf{min-max} & \textbf{min-max} &  &  \\
\midrule
\multirow{3}{*}{%
  \begin{tabular}{@{}c@{}}
  nl2postcond \\
  (Baseline)
  \end{tabular}
} 
  & \multirow{3}{*}{Single-pass} 
  & \multirow{3}{*}{1.0} 
  & \multirow{3}{*}{1.0} 
  & \multirow{3}{*}{46.0\%} 
  & \multirow{3}{*}{12.4\%} \\
& & & & & \\
& & & & & \\
\hline
\hline
R.Sampl. w. nl2postcond & \multirow{3}{*}{%
  \begin{tabular}{@{}c@{}}
  $\tau$ = 50 \\
  $\mu$ = 4
  \end{tabular}
} & 3.0 (1 - 4) & 3.0 (1 - 4) & 69.8\% & 25.13\% \\
Greedy & & 2.3 (1 - 4) & 2.3 (1 - 4) & 84.74\% & 65.14\% \\
Exploratory & & 3.8 (3 - 4) & 1.0 (1 - 2) & 92.2\% & 81.68\% \\
\hline
R.Sampl. w. nl2postcond & \multirow{3}{*}{%
  \begin{tabular}{@{}c@{}}
  $\tau$ = 50 \\
  $\mu$ = 8
  \end{tabular}
} & 5.1 (1 - 8) & 5.1 (1 - 8) & 77.97\% & 31.24\% \\
Greedy & & 2.9 (1 - 8) & 2.9 (1 - 8) & 88.05\% & 66.48\% \\
Exploratory & & 5.7 (3 - 8) & 1.2 (1 - 3) & 97.48\% & 85.3\% \\
\hline
R.Sampl. w. nl2postcond& \multirow{3}{*}{%
  \begin{tabular}{@{}c@{}}
  $\tau$ = 50 \\
  $\mu$ = 12
  \end{tabular}
} & 6.9 (1 - 12) & 6.9 (1 - 12) & 80.62\% & 40.59\% \\
Greedy & & 3.3 (1 - 12) & 3.3 (1 - 12) & 90.21\% & 69.94\% \\
Exploratory & & 6.5 (3 - 12) & 1.3 (1 - 5) & 98.43\% & 83.21\% \\
\hline
\hline
R.Sampl. w. nl2postcond & \multirow{3}{*}{%
  \begin{tabular}{@{}c@{}}
  $\tau$ = 70 \\
  $\mu$ = 4
  \end{tabular}
} & 3.5 (1 - 4) & 3.5 (1 - 4) & 69.84\% & 25.13\% \\
Greedy & & 2.6 (1 - 4) & 2.6 (1 - 4) & 86.16\% & 67.34\% \\
Exploratory & & 3.8 (3 - 4) & 1.0 (1 - 2) & 97.48\% & 82.75\% \\
\hline
\end{tabular}

\end{minipage}
\hfill
\begin{minipage}{0.49\textwidth}

\begin{tabular}{|l|c|c|c|c|c|}
\toprule
\textbf{Method} & \textbf{Config.} & \textbf{Attempts} & \textbf{Avg Subs} & \textbf{Corr.} & \textbf{Comp.} \\
 & & \textbf{min-max} & \textbf{min-max} & & \\
\midrule
R.Sampl. w. nl2postcond& \multirow{3}{*}{%
  \begin{tabular}{@{}c@{}}
  $\tau$ = 70 \\
  $\mu$ = 8
  \end{tabular}
} & 6.4 (1 - 8) & 6.4 (1 - 8)  & 77.97\% & 31.24\% \\
Greedy & & 3.8 (1 - 8) & 3.8 (1 - 8) & 91.82\% & 70.55\% \\
Exploratory & & 5.7 (2 - 8) & 1.2 (1 - 4) & 98.11\% & 85.46\% \\
\hline
R.Sampl. w. nl2postcond & \multirow{3}{*}{%
  \begin{tabular}{@{}c@{}}
  $\tau$ = 70 \\
  $\mu$ = 12
  \end{tabular}
} & 9.0 (1 - 12) & 9.0 (1 - 12) & 80.62\% & 40.59\% \\
Greedy & & 5.3 (1 - 12) & 5.3 (1 - 12) & 93.26\% & 74.51\% \\
Exploratory & & 6.5 (3 - 12) & 1.4 (1 - 6) & 99.12\% & 84.31\% \\
\hline
\hline
R.Sampl. w. nl2postcond & \multirow{3}{*}{%
  \begin{tabular}{@{}c@{}}
  $\tau$ = 90 \\
  $\mu$ = 4
  \end{tabular}
} & 3.7 (1 - 4) & 3.7 (1 - 4) & 69.84\% & 25.13\% \\
Greedy & & 3.2 (1 - 4) & 3.2 (1 - 4) & 86.79\% & 67.65\% \\
Exploratory & & 3.8 (3 - 4) & 1.1 (1 - 2) & 93.08\% & 80.79\% \\
\hline
R.Sampl. w. nl2postcond & \multirow{3}{*}{%
  \begin{tabular}{@{}c@{}}
  $\tau$ = 90 \\
  $\mu$ = 8
  \end{tabular}
} & 7.1 (1 - 8) & 7.1 (1 - 8) & 77.97\% & 31.24\% \\
Greedy & & 5.1 (1 - 8) & 5.1 (1 - 8) & 86.16\% & 69.68\% \\
Exploratory & & 6.2 (3 - 8) & 1.4 (1 - 4) & 98.74\% & 84.93\% \\
\hline
R.Sampl. w. nl2postcond & \multirow{3}{*}{%
  \begin{tabular}{@{}c@{}}
  $\tau$ = 90 \\
  $\mu$ = 12
  \end{tabular}
} & 10.3 (1 - 12) & 10.3 (1 - 12) & 80.62\% & 40.59\% \\
Greedy & & 6.2 (1 - 12) & 6.2 (1 - 12) & 87.42\% & 69.41\% \\
Exploratory & & 7.2 (3 - 12) & 1.7 (1 - 6) & {\bf 99.37\%} & {\bf 87.57\%} \\
\bottomrule
\end{tabular}
\end{minipage}
\end{table*}

{\color{custom-blue}{Table~\ref{tab:rq1-glm} shows the results as we used GLM-4.7 as the underlying model. As seen, Exploratory also achieves peak performance at $\tau$=90, $\mu$=12 ({\bf 99.37\% correctness, 87.57\% completeness}) with the same trend as the results using Llama Scout in Table~\ref{tab:rq1}.}}

\subsection{Related Work}
\label{sec:related}

Researchers have proposed automated approaches that fall into three categories: {\em program analysis-based}, {\em data mining-based}, and {\em large language model (LLM)-based}.

Program analysis-based methods use dynamic or static techniques. Dynamic approaches~\cite{ammons-popl02,BeschastnikhBSSE2011,daikon99} infer properties by monitoring executions but depend heavily on test coverage. Static analyses~\cite{10.1145/1273463.1273487,engler-sosp01,kremenek06,ramanathan-pldi07,yiwei-icse11} and abstract interpretation~\cite{10.1145/2384616.2384633} avoid this dependency but often produce conservative or imprecise specifications due to high false-positive rates. Several approaches attempted to generate specification by analyzing API documentation or code comments~\cite{10.5555/2337223.2337319,10.1145/1294261.1294276,10.1145/1985793.1985796,10.1109/ICST.2012.106,10.1145/3213846.3213872,10.1109/ASE.2009.94,7985647}.

Data mining-based approaches extract common API usage patterns from large codebases~\cite{prminer-fse05,dynamine05,mapo09}. They identify call pairs, sequences, and finite-state models~\cite{zeller07,williams-tse05,taoxie-ase09,fse09,mike-ase09,zeller-ase09}. While useful for API usages, they rarely infer semantic specifications such as pre/postconditions~\cite{ramanathan-pldi07}.


In general, ML approaches for specification generation have advanced in multiple directions, including postcondition inference~\cite{nl2postcond}, synthesizing test oracles~\cite{dinella2022toga,mastropaolo2023usingtransfer,10.1145/3524481.3527220}, improving test coverage~\cite{lemieux2023codamosa,coverup2025, le2026testweaver}, and generating unit tests~\cite{lahiri2023interactivecodegenerationtestdriven,tufano2021unittestcasegeneration}. These techniques rely on different types of inputs. AthenaTest~\cite{tufano2021unittestcasegeneration} generates both the input and the oracle of a unit test directly from the implementation of the focal method, while TOGA generates only the oracle~\cite{dinella2022toga}. TiCoder~\cite{lahiri2023interactivecodegenerationtestdriven} leverages LLMs to produce both inputs and outputs based on a natural language description of user intent.

\begin{figure}[t]
    \centering
    \includegraphics[width=0.5\textwidth]{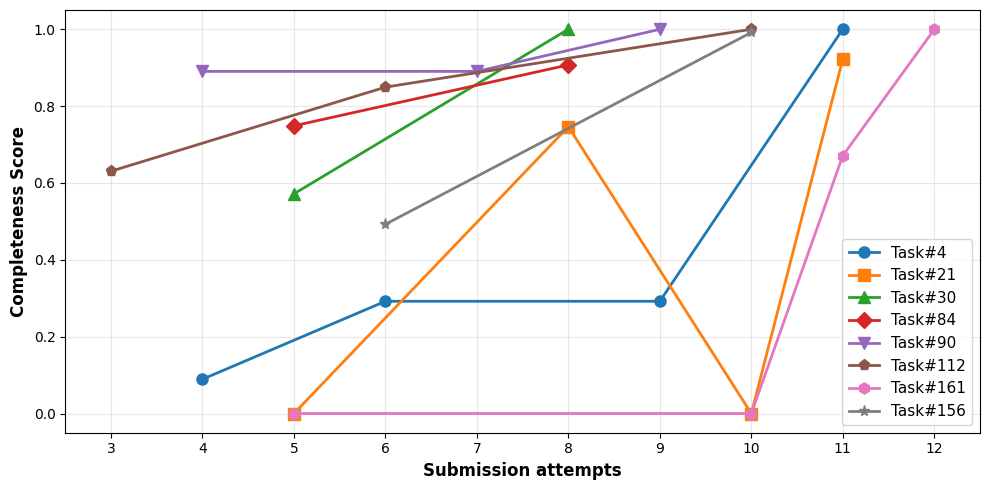}
    \vspace{-21pt}
    \caption{Trend of completeness score from our seleted cases in \textit{EvalPlus}, using Exploratory Multi-turn with configuration $\tau$=90 and $\mu$=12}
    \label{fig:trend}
\end{figure}

While these methods target the production of concrete test cases (and sometimes oracles), our work instead focuses on generating postconditions. Similarly, EvoSpex~\cite{10.1109/ICSE43902.2021.00112} applies evolutionary learning to infer functional input–output relationships.
More recent efforts have explored property-based specifications. Vikram {\em et al.}~\cite{vikram2024largelanguagemodelswrite} propose leveraging LLMs to generate property-based tests (PBTs), while Speculyzer\cite{DBLP:journals/corr/abs-2210-00848} enumerates likely properties and candidate inputs with LLMs, but uses them primarily as heuristics to guide code generation rather than to ensure correctness or completeness of specifications. Beyond input–output properties, machine learning has also been applied to infer intermediate specifications, such as program invariants~\cite{10.1145/2837614.2837664,Laich2020Guiding,10.1145/3385412.3385986,pmlr-v202-pei23a}.


\subsection{\bf Behavioral Patterns in Completeness Across Submissions}
\label{sec:appendix_patter}


\begin{table*}[ht]
\centering
\scriptsize
\caption{Exploratory Multi-turn's trajectory for function \#159 in EvalPlus dataset. It requires implementing the function \code{eat}, which calculates the total number of carrots eaten and the number of carrots remaining, given \code{number} (carrots already eaten), \code{need} (carrots that need to be eaten), and \code{remaining} (carrots in stock).}
\label{tab:reasoning_categories}
\vspace{-6pt}
\tabcolsep 2pt
\begin{tabular}{c@{\hskip 8pt}p{2.4cm}@{\hskip 8pt}p{10.7cm}}
\toprule
\textbf{Att.} & \textbf{Category} & \textbf{Reasoning (ret=return\_value, rem=remaining, num=number.)} \\
\midrule
1 & {\em Verify Func. Behavior} & The function `eat` calculates... The function returns... Key constraints and edge cases:... To verify the correctness of the function, we need to write symbolic postconditions that check these properties. \\ 
 & & \texttt{\textcolor{blue}{>> assert isinstance(ret, list) \&\& len(ret) == 2 \&\& isinstance(ret[0], int) \&\& isinstance(ret[1], int)}} \\
2 & {\em Check Edge Cases} & The first assertion checks... However, we need to add more assertions to cover other properties of the function. Let's consider the case when `need` is less than or equal to `remaining`... We should also consider the case when `need` is greater than `remaining`... \\
 & & \texttt{\textcolor{blue}{>> assert need <= rem || (need > rem \&\& ret[1] == 0)}} \\
3 & {\em Refine Assertions} & The second assertion checks if... should also check if the total number of carrots eaten is correct in both... \\
 & & \texttt{\textcolor{blue}{>> assert (need <= rem \&\& ret[0] == num + need) || (need > rem \&\& ret[0] == num + rem)}} \\
4 & {\em Postcondition Combining} & ... let's try to combine them into a single assertion... \\
& & \texttt{\textcolor{blue}{>> assert (need <= rem \&\& ret[1] == rem - need) || (need > rem \&\& ret[1] == 0)}} \\
& & \texttt{\textcolor{blue}{>> assert (need <= rem \&\& ret[0] == num + need) || (need > rem \&\& ret[0] == num + rem)}} \\
5 & {\em Refine Assertions} & ... let's refine our assertions to make them more concise and meaningful... \\
& & \texttt{\textcolor{blue}{>> assert (need <= rem \&\& ret[0] == num + need \&\& ret[1] == rem - need) || (need > rem \&\& ret[0] == num + rem \&\& ret[1] == 0)}} \\
6 & {\em Submit Final Solution} & We have several assertions that check different properties of the function... Let's try to finalize our solution by submitting a single assertion that combines all the checks... \\
& & \texttt{\textcolor{blue}{>> assert (need <= rem \&\& ret == [num + need, rem - need]) || (need > rem \&\& ret == [num + rem, 0])}}\\
\bottomrule
\end{tabular}
\label{tab:rq2_sample}
\end{table*}

To analyze the change in completeness score across submission attempts, we consider the strictest setting with completeness threshold $\tau=90$ and maximum number of turns $\mu=12$. We plot the completeness score over submission attempts for eight selected solvable cases (i.e., cases where the final solution is correct and satisfies the completeness target) in Fig.~\ref{fig:trend}. These cases are representative of typical behaviors observed in our dataset and are selected here for illustration.

The selected tasks require between two and four submission attempts. All eight cases show an increase in completeness from the first submission to the final submission that reaches the completeness threshold. Across all 164 tasks, we observe that 23 cases show no change in completeness between consecutive submissions (e.g., case \#90 and case \#161), indicating early convergence. In contrast, 18 cases temporarily drop to zero completeness due to the submission of an incorrect postcondition before recovering in later attempts (e.g., case \#10, not shown). The earliest correct submission occurs at attempt 3 in 11 cases (e.g., case \#112), suggesting that Exploratory Multi-turn often identifies a correct core postcondition early and incrementally improves its discriminative power.
\vspace{-4pt}
\subsection{\bf A Case Study of Exploratory vs. Greedy Multi-turn}
\label{sec:appendix_case}
\begin{figure*}[t]
    \centering
    \includegraphics[width=\textwidth]{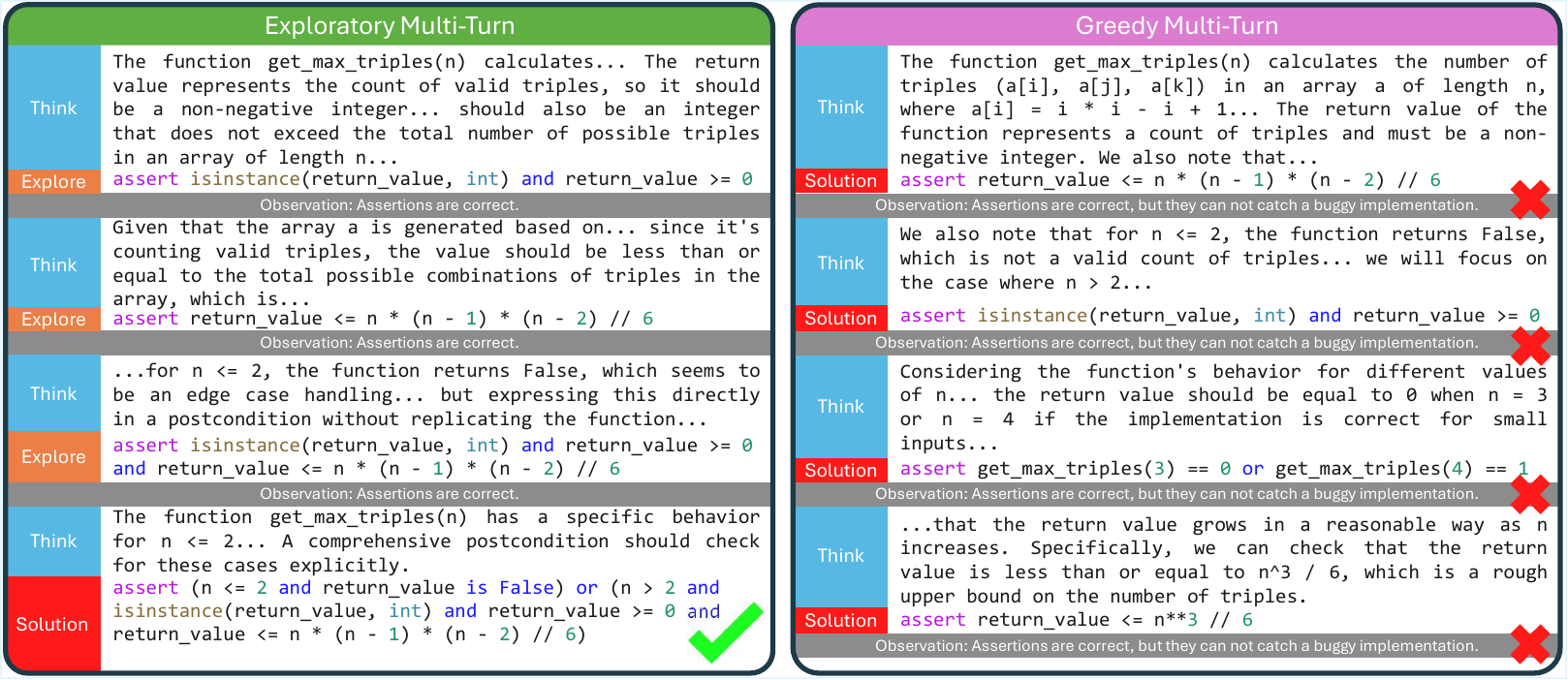}
    \caption{
A comparison between Exploratory Multi-turn and Greedy Multi-turn for case \#147 from \textit{EvalPlus}: the figure illustrates {\tool}'s outputs ($\tau$=50 and $\mu$=4), model's reasoning are given in the blue blocks, submission attempts are given in the red blocks, and exploration attempts are enclosed in the orange blocks. The feedback is shown in the gray boxes. {\textcolor{green}{\checkmark}}: correct and complete postcondition, {\textcolor{red}{$\times$}}: otherwise.
}
    \label{fig:sample}
\end{figure*}

To illustrate the behavioral differences between Greedy and Exploratory Multi-turn, Fig.~\ref{fig:sample} presents a detailed analysis of case \#147 from EvalPlus under $\tau=50$ and $\mu=4$. The task defines the function \code{get\_max\_triples(n)}, which computes the number of valid triples in an array $a$ of length $n$, where $a[i] = i \times i - i + 1$, such that the sum of the triples is divisible by 3. If $n \leq 2$, the function returns \code{False}.

Exploratory Multi-turn submits its final solution at attempt \#4, whereas Greedy Multi-turn submits a candidate postcondition at every attempt by design, with its best-performing one appearing in attempt \#1. In the first attempt, both approaches identify two key properties: (1) the return value should be a non-negative integer, and (2) it should be bounded by the total number of possible triples. Exploratory probes the first property through an exploratory assertion, while Greedy directly submits the second, more complex constraint.

In the second attempt, Exploratory explores the second property and explicitly notes its potential relevance. Greedy, however, reverts to the first property and briefly mentions the case $n \leq 2$ in its reasoning but dismisses it as irrelevant, despite its semantic importance.

In the third attempt, Exploratory combines the two properties and explicitly incorporates the $n \leq 2$ case. In contrast, Greedy submits assertions targeting specific input values (e.g., $n=2$, $n=3$). Notably, its reasoning states \code{get\_max\_triples(n)=0}, while the submitted assertion enforces \code{get\_max\_triples(n)=1}; this mismatch happens to pass tests but reflects unstable reasoning.

In the final attempt, Exploratory refines its earlier reasoning by adding the missing $n \leq 2$ sub-condition and submits a complete postcondition. Greedy instead submits a new constraint, \code{return\_value <= n**3 // 6}, which is weaker than a stricter constraint it had already produced earlier, resulting in a correct yet incomplete specification.

Overall, Exploratory Multi-turn produces a correct and complete postcondition using a single submission and three exploratory steps. It systematically explores sub-conditions and composes them into a final specification. In contrast, Greedy Multi-turn identifies a partially correct condition early but fails to refine it into a comprehensive postcondition within the same budget.

\subsection{\bf An Example Trajectory of Reasoning Categories}
\label{sec:appendix_reasoning_example}


Table~\ref{tab:rq2_sample} presents a representative reasoning trajectory produced by Exploratory Multi-turn for case \#159 from the EvalPlus dataset. This task requires implementing the function \code{eat}, which computes the total number of carrots eaten and the number of carrots remaining, given \code{number} (already eaten carrots), \code{need} (carrots to be eaten), and \code{remaining} (carrots in stock).

The trajectory begins with {\em Verify Function Behavior}, where the model describes the function’s overall behavior and proposes assertions checking basic structural properties of the return value. It transitions to {\em Check Edge Cases}, explicitly reasoning about conditional scenarios such as whether \code{need} exceeds \code{remaining}. The model enters {\em Refine Assertions}, improving earlier checks by strengthening semantic constraints on the return values.

The reasoning then moves to {\em Postcondition Combination}, where  previously explored assertions are consolidated. This is followed by another {\em Refine Assertions} step, reflecting iterative improvement toward a more concise and comprehensive specification. Finally, the trajectory concludes with {\em Submit Final Solution}, where the model submits a single assertion that captures all explored conditions.

Overall, this illustrates a systematic progression through reasoning categories, starting from high-level function understanding, moving to edge case analysis and iterative refinement, and culminating in the submission of a complete postcondition.

\subsection{An Example of Bugs Detected by Conditions generated by {\tool}}
\label{appendix:bug-example}

\vspace{2pt}
Let us use an example of the correct code (not shown), aiming to solve the task: {\em given a starting point and a mapping of each point to the next, find the position reached after $K$ steps.}  




1. When running the baseline \code{nl2postcond} in Random Sampling $\mu$ independent times, the LLM does not learn from its past experience, leading to repeated mistakes. For example, the postconditions produced at the attempts \#2 and \#5 are the same even though the one at attempt \#2 is incorrect. This approach exhausted all 10 attempts without catching the bug.




2. For Greedy approach, the LLM demonstrated the ability to learn from past mistakes. For example, in attempt \#2, it proposed to check if the output is a positive number:

{\color{orange}{
{\tiny\ttfamily
\code{assert isinstance(return\_values[0], int) and return\_values[0] > 0}
}}}

Getting the correctness feedback, it simplified the condition to only check if the output is an integer: \code{assert isinstance(return\_values[0], int)}. However, none of the 10 retries can catch the bug.






3. Exploratory Multi-turn started with a simple postcondition checking if the output is an integer. After getting the feedback on its incorrect condition, it {\bf expanded the condition to a correct one} to check that the output string can be converted to an integer.

We also observed that Exploratory was able to {\bf refine the postcondition}. For example, at attempt \#3, it checks that the output index is within the bounds of the sequence:

{\color{orange}{
\code{N, K = map(int, lines[0].split());} \code{sequence = list(map(int, lines[1].split()));}

\code{assert 1 <= int(return\_values[0]) <= N}
}}

After getting the positive feedback, it continued to refine the post condition to a stricter one and eventually submitted the correct one at the attempt \#5. Its reasoning texts are as follows: {\em "The current postcondition checks if the output index is within the bounds ... However, to make the postcondition more specific and meaningful, I will consider...Given that the sequence represents a linked list with possible cycles, the output value should be an element in the sequence."} The final condition caught the buggy version as the output is not an element of the sequence.

{\color{orange}{
\code{N, K = map(int, lines[0].split());} \code{sequence = list(map(int, lines[1].split()))}

\code{assert int(return\_values[0]) in sequence}
}}

\subsection{Usage of LLM Assistance}  
We used large language models (LLMs) to aid in polishing the writing and improving clarity. All research ideas, experiments, and conclusions are the work of the authors. 

\end{document}